\documentclass[pra,showpacs,twocolumn,superscriptaddress,floatfix,aps]{revtex4-2}
\usepackage{amsmath,amssymb,bm}
\usepackage{graphicx}
\usepackage{physics}
\usepackage[T1]{fontenc}
\usepackage{comment}
\usepackage{float}
\usepackage{mathtools}
\usepackage[normalem]{ulem}
\usepackage{hyperref}
\usepackage{soul}
\usepackage{braket}
\usepackage[utf8]{inputenc}
\hypersetup{colorlinks,%,%
 linkcolor=blue,%
 citecolor=blue,%
 urlcolor=blue}
\usepackage{color}
\usepackage{xcolor}
\usepackage[dvipsnames]{xcolor}
\usepackage{tikz}
\usetikzlibrary{shapes}
\usepackage{placeins}

\usepackage{soul,xcolor} %Strike through this text

\def\dar{\downarrow}
\def\upar{\uparrow}

\begin{document}
\setstcolor{red}

\title{Rabi-induced localization and resonant delocalization of a binary condensate in a spin-asymmetric quasiperiodic potential}

\author{Swarup K. Sarkar}
%\email{skanti@iitg.ac.in}
\affiliation{Department of Physics, Indian Institute of Technology Guwahati, Guwahati 781039, Assam, India}

\author{Sh. Mardonov}
\affiliation{New Uzbekistan University, Movarounnahr str. 1, Tashkent 100000, Uzbekistan}
\affiliation{Ulugh Beg Astronomical Institute, Tashkent, 100052, Uzbekistan}

\author{E. Ya. Sherman}
\affiliation{Department of Physical Chemistry, University of the Basque Country UPV/EHU, 48940 Leioa, Spain}
\affiliation{IKERBASQUE, Basque Foundation for Science, Bilbao, Spain}
\affiliation{EHU Quantum Center, University of the Basque Country UPV/EHU, 48940 Leioa, Spain}
% \email{Corresponding author, evgeny.sherman@ehu.eus}
%\footnote{Corresponding author}

\author{Pankaj K. Mishra}
%\email{pankaj.mishra@iitg.ac.in}
\affiliation{Department of Physics, Indian Institute of Technology Guwahati, Guwahati 781039, Assam, India}
\date{\today}

% \date{\today}

\begin{abstract}
We theoretically investigate the ground state and dynamics of a Rabi-coupled pseudospin-1/2 Bose-Einstein condensate, where only one spin component is subjected to an external potential. We show that in the quasiperiodic potential the Rabi coupling induces localization between the components as it is raised above the threshold value. Interestingly, the localization is {\it mutually induced} by both components for the quasiperiodic confinement, whereas for a harmonic trap the localization is induced in the potential-free component by interaction with that confined in the potential. Further, we explore the condensate dynamics by implementing a periodic driving of the Rabi frequency, where various frequency-dependent delocalization patterns,  such as double (triple)-minima, tree-(parquet)-like, and frozen distributions with a correlated propagation of different spin populations are observed in the condensate density. These features pave the way to control the condensate mass and spin density patterns, both in the stationary and dynamical realizations. 

\end{abstract}

\date{\today}

\maketitle

\section{Introduction}
%%%%%%%%%%%%%%%%%%%%%%%%%%%
Bose-Einstein condensates (BECs), as macroscopic quantum states of matter, offer a highly controllable platform for exploring fundamental quantum phenomena, including the effects of disorder and interactions in low-dimensional systems. Among the most intriguing phenomena observed in such systems is the Anderson localization, originally proposed to describe the suppression of electronic diffusion in the disordered media~\cite{Anderson:1958}, since then it has become a typical model for understanding the localization in a wide variety of complex systems. These include photonic lattices~\cite{Wiersma:1997, Scheffold:1999, Schwartz:2007, Aegerter:07, Topolancik:2007}, microwaves~\cite{Dalichaouch:1991, Dembowski:1999, Chabanov:2000, Pradhan:2000}, acoustic waves~\cite{WEAVER:1990}, and ultracold atomic gases~\cite{Billy:2008, Roati:2008}. In ultracold atomic gases, the weakly interacting BECs have been a suitable platform for exploring the localization of the quantum matter waves. Following the first experimental realization of localization of the condensate in the random~\cite{Billy:2008} and quasiperiodic~\cite{Roati:2008} potentials, the field has witnessed a great number of theoretical and experimental works. Various analytical~\cite{Sanchez:2006,Sanchez:2007} and numerical~\cite{Adhikari:2009, Cheng_random:2010,Cheng:2011, Cheuk:2012,Zohra:2024} approaches based on the mean-field Gross-Pitaevskii (GP) equation have been used to investigate the complex interplay between disorder and interactions on the localization. 

Recently, the experimental realization of pseudospin-1/2 BECs where two hyperfine state are coupled with the synthetic spin-orbit (SO) and Rabi couplings generated using the Raman lasers~\cite{lin:2011} has become a fertile ground for investigating the rich physics of spin-dependent localization in condensates in quasiperiodic~\cite{cheng:2014,Li:2016,Sarkar:2024} and random potentials~\cite{Xi:2015,Zhang:2022, Sarkar:2025}. 
These systems also exhibit very intriguing nonlinear dynamics due to the spin-dependent velocities ~\cite{Mardonov2015, Mardonov2018, zhang:2023}. Additionally, it has been reported that the interplay between the inter- and intraspecies interactions, especially for a lifted Manakov's symmetry~\cite{Manakov:1974,Cheng_spat:2010}, {where the nonlinear contribution to the BEC energy depends on the spin orientation,} can produce a spatially asymmetric localization of the condensate~\cite{Sarkar:2025}. 

A comprehensive understanding of localization in SO and Rabi coupled BECs remains challenging due to the complex interplay between disorder, interactions, and spin-related couplings~\cite{Sarkar:2025}, especially if SO and Rabi coupling do not commute. While the SO coupling promotes a spatial decoupling of spin components, the Rabi interaction fosters their coupling. To address this complexity, various simplified models have been proposed in recent years. Those often involve asymmetric spin-dependent potentials, where one spin component experiences a trapping potential while the other does not. In such setups, the trapped component exhibits localization, which in turn induces localization in the untrapped one either through density-density interaction or through the Rabi coupling. In this context, Wang {\it et al.} ~\cite{wang:2019} investigated induced localization in an SO-coupled BEC confined in a double-well potential. They demonstrated that an imbalance between intra- and inter-species interactions can drive a transition from a spin-balanced phase to a spin-localized phase. Similarly, Santos and Cardoso~\cite{Santos:2021} explored the localization in binary BEC where one component is trapped in a quasiperiodic potential and coupled to the free component via linear Rabi coupling and reported the related induced localization. 

Experimentally, tuning SO coupling has posed a severe challenge to the researchers. Several studies have proposed controlling SO coupling through rapid modulation of laser intensities~\cite{Zhang:2013, Jim:2015}. In a similar vein, time-modulated Rabi frequencies have been used to realize quantum phases~\cite{Struck:2011}, artificial gauge fields~\cite{Meinert:2016}, matter-wave control~\cite{Abdullaev:2010}, and to probe Landau-Zener tunneling~\cite{Olson:2014, Gomez:2024}. For rapid modulation, Deconinck {\it et al.} \cite{Deconinck:2004} derived analytical solutions for linearly coupled Gross-Pitaevskii equations using a unitary transformation that absorbs the time-dependent Rabi frequency under Manakov's symmetry. Building on this, Nistazakis {\it et al.} \cite{Nistazakis:2008} numerically implemented Rabi switch to transfer nonlinear structures between components, with reduced efficiency observed when the symmetry is broken. More recently, Abdullaev {\it et al.}~\cite{Abdullaev:2018} investigated parametric resonances and Josephson-like oscillations in SO-coupled BECs under time-modulated Raman coupling.

As we have seen, Rabi coupling significantly influences the ground state and dynamics of binary condensates. Trombettoni {\it et al.} \cite{Trombettoni:2009} developed a theoretical framework showing that Rabi coupling enables population exchange in deep optical lattices. It also mediates the transformation of dark solitons into vector dark solitons\cite{Liu:2012}, and affects miscibility in both non-dipolar~\cite{Merhasin:2005, Merhasin:2005phy_script} and dipolar condensates~\cite{Gligori:2010, Chiquillo:2018}, where it is crucial for immiscibility–miscibility transitions~\cite{Gligori:2010}. Moreover, it facilitates complex excitations like vector rogue waves in multi-component BECs~\cite{Kanna:2019}, stabilizes soliton-like states under time-modulated coupling in quasi-2D systems~\cite{sushanto:2008}, and inhibits vortex formation in rotating spinor BECs~\cite{Zhao:2021rot_bec}.

Although the model analysis of induced localization by Santos and Cardoso~\cite{Santos:2021} provides a basic understanding of the roles played by Rabi coupling and interactions in the emergence of localized states, many fundamental aspects, both stationary and dynamical, of these states remain largely unexplored. In this work we study and understand these intriguing localizations  and delocalizations demonstrating a rich set of quantum mechanical effects.

This paper is organized as follows. In Sec.~\ref{sec2}, we formulate the mean-field model with binary Gross-Pitaevskii equations and main observables to characterize the system of interest. Section ~\ref{sec:3} presents numerical results by showing the effect of linear Rabi coupling in the ground state, including the critical behavior of the localization and a strong impact of even weak nonlinearities.  Subsequently, in Sec.~\ref{sec:4} we discuss the effect of time periodic Rabi frequency on the localized condensates, both linear and those with lifted Manakov's symmetry of nonlinearities. Finally, we conclude our work in Sec.~\ref{sec:conclusion} and present some details, including a comparison of induced localization in quasiperiodic and harmonic traps, in the Appendix.

%%%%%%%%%%%%%%%%%%%%%%%%%%%%%%%%%%%%%
\section{Mean-field model and  observables}
\label{sec2}

In this Section we discuss the mean-field dynamical model used in the present work and define the relevant observables used to characterize the induced localizations.

\subsection{Coupled Gross-Pitaevskii equations}
We consider a pseudospin-$1/2$ quasi-one-dimensional condensate trapped strongly in the transverse direction, modeled by the coupled Gross-Pitaevskii equations~\cite{Santos:2021,RAVISANKAR:2021,Sarkar:2025}:%
\begin{subequations}%
\label{eqn1}%
\begin{align}%
% \begin{split}
 {\mathrm i} \frac{\partial \psi_{\upar}}{\partial t} = & \left[-\frac{1}{2}\frac{\partial^{2}}{\partial x^{2}} + 
 g_{\upar \upar} \vert \psi_{\upar} \vert ^{2} + 
 g_{\upar\dar} \vert \psi_{\dar} \vert ^{2} + V_{\upar}(x)\right]\psi_{\upar} \notag \\
 & + \Omega_{0} \psi_{\dar}, \label{eqn1(a)}\\ 
 {\mathrm i} \frac{\partial \psi_{\dar}}{\partial t} = & \left[-\frac{1}{2}\frac{\partial^{2}}{\partial x^{2}} + 
 g_{\dar\dar} \vert \psi_{\dar} \vert ^{2} + g_{\dar \upar} \vert \psi_{\upar}\vert^{2} + V_{\dar}(x)\right]\psi_{\dar} \notag \\ &
 + \Omega_{0} \psi_{\upar}, \label{eqn1(b)} 
\end{align}
\end{subequations}%
where $\psi_{\upar}$ and $\psi_{\dar}$ $(\psi_{\upar,\dar}\equiv \psi_{\upar,\dar}(x,t))$ represent the pseudo spin-up and spin-down components of the condensate wavefunction ${\bm\psi}=\left(\psi_{\upar},\psi_{\dar}\right)^{\rm T}$, respectively, where ${\rm T}$ stands for transposition. For stationary states $\psi_{\upar,\dar}(x,t)=\psi_{\upar,\dar}(x)\exp(-{\mathrm i}\mu t),$ where $\mu$ is the chemical potential. 
 Here, $g_{\upar\upar}$ and $g_{\dar\dar}$ are the intra-species interaction strengths, $g_{\upar\dar}$, and $g_{\dar\upar}$ are inter-species interactions {(with $g_{\upar\upar}=g_{\dar\dar}=g_{\upar\dar}=g_{\dar\upar}$ corresponding to the Manakov's symmetry \cite{Manakov:1974})}, and $\Omega_{0}$  is the Rabi coupling strength. 
From now on for brevity, we will remove the $(x,t)$ notations in  $\psi_{\upar (\dar)}$  when it does not cause confusion.

The GPEs (\ref{eqn1(a)})-(\ref{eqn1(b)}) correspond to the binary BEC Hamiltonian with the linear Rabi coupling $\Omega_{0}\sigma_{x}$ with $\sigma_{x} $ being the corresponding Pauli matrix. The trapping potential is $V_{j}(x)$ with $j = \{\upar, \dar\}$. Here we consider the condensate interacting with the trapping potential only in the spin-up component~\cite{Santos:2021}, such as: 
\begin{align}\label{eq:VupVdown}
V_{\upar}(x) =  V(x), \qquad
V_{\dar}(x)  = 0. 
\end{align}
{In general, this spin asymmetry makes the BEC localization features qualitatively different from those expected for the $V_{\dar}(x)=V_{\upar}(x) =  V(x)$ potential.} In order to analyze the localization induced  by one spin component on the other we consider $V(x)$ as a quasiperiodic potential of the form:
\begin{align}
V(x) =  V_1 \sin^2(k_1 x) + V_2 \sin^2(k_2 x), \label{eqn3} 
\end{align}
where $V_1$ and $V_2$ are the primary and secondary optical lattice amplitudes, respectively. This potential has minima at points $x=x_{i}\geq 0$ and $x=-x_{i} <0$ ($i=0,1\ldots,$ $x_{0}=0,$ and $x_{i+1}>x_{i}$) where 
\begin{eqnarray}\label{eq:Vprimes}
&&V^{\prime}(x) =  V_1 k_1\sin(2k_1 x_{i}) + V_2 k_2 \sin(2k_2 x_{i})=0, \\  
\notag \\
&&V^{\prime\prime}(x) =  2\left(V_1 k_1^{2}\cos(2k_1 x_{i}) + V_2 k_2^{2}\cos(2k_2 x_{i})\right)>0,    \notag
\end{eqnarray}
characterized by corresponding local oscillator frequencies $\omega_{i}=\sqrt{V^{\prime\prime}(x)}$ at $x=x_{i}.$

In experiments, the pseudospin-1/2 Rabi-coupled BEC can be realized by using two hyperfine states of $^{87}$Rb atoms as pseudo spin-up $\ket{\upar} \equiv \ket{F = 1, m_F = 0}$ and spin-down $\ket{\dar} \equiv \ket{F = 1, m_F = -1}$ which are coupled by a pair of Raman lasers with wavelength $804.1$ nm. Recoil momentum $k_L = \sin(\theta/2)k_{p},$ where $k_{p}$ is the photon wavevector, and the recoil energy $E_L = \hbar^2k_L^2/2m$  provide relevant scales for tuning the SO and Rabi couplings ~\cite{Spielman:2011, Jim:2015}. In quasi-1D, the condensate is strongly confined in transverse direction with $\omega_{\perp} \sim 10^{3}$ s$^{-1}$ with corresponding $a_{\perp} \sim 1\ \mu$m and $\hbar\omega_{\perp} \sim 5$ nK. The inter and intra-spin scattering lengths being typically of the order of 5 nm  can be further controlled by using Feshbach resonances.

To obtain the dimensionless Eq.~(\ref{eqn1(a)})-(\ref{eqn1(b)}), we consider the transverse harmonic oscillator length $a_{\perp} = \sqrt{\hbar/(m\omega_{\perp})}$ as a characteristic length scale with $\omega_{\perp}$ as the transverse harmonic trapping frequency, $\omega_{\perp}^{-1}$ as the timescale and $\hbar \omega_{\perp}$ as the characteristic energy scale. The interaction parameters can be defined in terms of {$g_{{\upar\upar},({\dar\dar})}= 2\mathcal{N}a_{{\upar\upar},({\dar\dar})}/a_{\perp}$,} and $g_{\upar\dar} = 2\mathcal{N}a_{\upar\dar}/a_{\perp}$, where, $a_{{\upar\upar},({\dar\dar})}$ and $a_{\upar\dar}$ represent the intra- and inter-component scattering lengths, respectively, and $\mathcal{N}$ represents the total number of atoms in the condensate. The dimensionless Rabi coupling $\Omega_{0}$ is defined in the units of $2\omega_{\perp}$ with the wavefunction being rescaled with $\sqrt{a_{\perp}}.$ 

\subsection{Definition of spin-dependent observables} \label{sec:IIB}

To characterize the localization and delocalization at different spatial scales we start with the occupation numbers $N_{j},$ where
\begin{equation}
N_{j}=\int_{-\infty}^{\infty} \vert \psi_{j} \vert ^{2} dx,
 \label{eqn:N}
\end{equation} 
with $N_{\upar}+N_{\dar}=1.$ The width $w_{j}$ is given by
\begin{align}
\label{eq:wj2}
w_{j}^{2} = \frac{1}{N_{j}}\int_{-\infty}^{\infty} (x-\langle x_{j} \rangle)^{2} \vert \psi_{j} \vert ^{2} dx, 
\end{align}
where the center of mass position $\langle x_{j}\rangle:$ 
\begin{align}\label{eq:xm}
\langle x_{j} \rangle = \frac{1}{N_{j}}
\int_{-\infty}^{\infty} x \vert \psi_{j} \vert ^{2} dx. 
\end{align} 
The shape is characterized by the inverse participation ratio (IPR) $\chi_{j}:$
\begin{align}\label{eq:chij}
\chi_{j}=\frac{1}{N_{j}^{2}}\int_{-\infty}^{\infty} \vert \psi_{j} \vert ^4 dx.
\end{align}

We also utilize the ``spin miscibility'' parameter characterizing the joint distribution of densities of spin components defined as
\begin{align} \label{eq:misc}
\eta=2 \int_{-\infty}^{\infty} \vert {\psi}_{\upar}\vert \vert {\psi}_{\dar} \vert dx,
\end{align}
with $\eta=1$ ($\eta=0$) corresponding to fully miscible (immiscible) realizations. For real wavefunctions $\eta=\vert \langle\sigma_{x}\rangle\vert $.

After obtaining the initial ground state $\psi_{j}(x,0)$, we use the time-varying Rabi frequency to study the dynamics of the condensate. For this purpose we calculate the time correlation function corresponding to different states which is defined in terms of the absolute value of the overlap function as,
\begin{align}
C_{j}(t) = \Big\lvert \int_{-\infty}^{\infty} \psi_{j}(x,t) \psi_{j}(x,0) dx \Big \rvert, 
\label{eqn:c_t}
\end{align}
where $C_{j}(0) = N_{j}(0).$  As we will see below, the criteria of induced delocalization can be obtained by considering evolution of $C_{j}(t)$ for given spin state~\cite{Sarkar:2023}.  

\section{Ground state of induced localization}
\label{sec:3}
To understand the role of Rabi coupling in inducing localization and shaping the ground-state structure of a binary condensate, we present a detailed analysis of the ground-state profiles of the spin components when they are linearly coupled via the Rabi frequency. We begin by examining the non-interacting case and investigate how Rabi coupling influences localization in the spin-down component induced by the spin-up component, and vice versa. The results of imaginary time propagation are complemented by an eigenmode analysis, which allows us to determine the threshold Rabi frequency beyond which induced localization emerges in the spin components. We then extend this analysis to include interactions, exploring how Rabi coupling affects induced localization in the interacting binary condensate.

\subsection{Calculation procedure}

We begin our analysis by presenting the numerical results obtained by solving the pair of coupled GPEs (Eqs.~(\ref{eqn1(a)})-(\ref{eqn1(b)})) in which the spin-up component is trapped with a bichromatic lattice potential (\ref{eqn3}) and the trapless spin-down component interacts with spin-up component by linear Rabi coupling. For all of our calculations, the primary and secondary lattice strengths are considered as $V_1 = 1$ and $V_2 = 0.5$, respectively, with wavenumbers $k_1 = 0.35$ and $k_2/k_1 = (\sqrt{5} - 1)/2,$  corresponding to the inverse golden ratio. 

To obtain the ground state we begin with the imaginary time propagation (ITP) considering the coupled GPEs (\ref{eqn1}) with the potential \eqref{eqn3} using time-splitting Fourier spectral method~\cite{jha:2023qutarang}. However, since the spin-down component is not subjected to the external potential, obtaining the ground state using the ITP becomes particularly challenging for low Rabi coupling strengths ($\Omega_0 < 0.2$). To address this, we also solve the linearized GPEs as an eigenvalue problem to compute the full spectrum of eigenvalues and corresponding eigenstates without any self-interactions. 

This matrix method involves constructing a $(2N \times 2N)$ matrix that represents the coupled linear GPEs, where $N$ is the number of grid points for the spatial domain being discretized at $x = [-L, L]$ with $L = N \Delta x/2.$ We use the same spatial step size $\Delta x = 0.025$ and grid size $N = 8192$ as in the ITP to ensure consistency. Once the matrix is constructed, it is diagonalized using the ARPACK package in Python. A key advantage of this matrix-based method is that it treats the problem as stationary, making it well-suited for exploring regimes with weak Rabi coupling. This allows us to probe the low $\Omega_0$ region more effectively than with the ITP approach.

Thus, to obtain the ground state for non-interacting part of the problem we use both the ITP and matrix method to solve the coupled GPEs, while, at nonzero self-interaction, we resort to the ITP only, by choosing the initial state as an antisymmetric Gaussian wavefunction: $\psi_{\upar}(x) = - \psi_{\dar}(x).$ The imaginary and real-time propagation is utilized with time step $\Delta t = 10^{-4}$ to study the ground state and dynamics of the condensate.

\subsection{Ground state of non-interacting BEC: Rabi-induced  localization}
\begin{figure}[!htp]
    \centering
    \includegraphics[width=\linewidth]{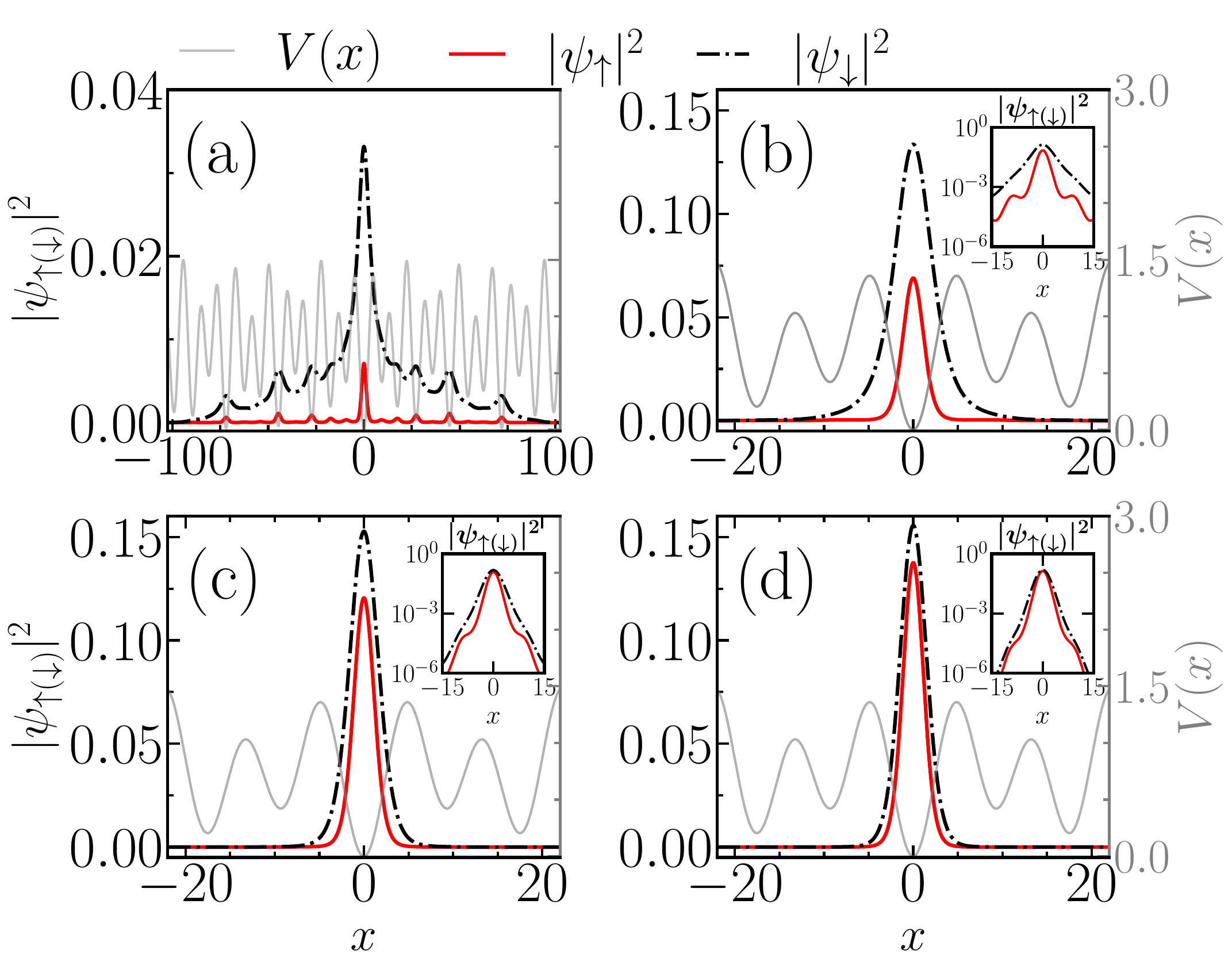}
    \caption{The BEC density $|\psi_{\upar(\dar)}|^2$ for different values of Rabi coupling $\Omega_0$: (a) $\Omega_0 = 0.08$, (b) $\Omega_0 = 0.2$, (c) $\Omega_0 = 0.4$, and (d) $\Omega_0 = 0.6.$ Increasing $\Omega_0$ equalizes the spin-up and spin-down  profiles.  In the insets of (b)-(d), the density is plotted on a semilogarithmic scale to highlight the exponential and Gaussian behavior of the localized condensate, as it depends on $\Omega_0$. Here, all the interactions $g_{\upar\upar} = g_{\dar\dar} = 0$.}
    \label{fig:density}
\end{figure}

\begin{figure}[!htp]
    \centering
   \includegraphics[width=0.85\linewidth]{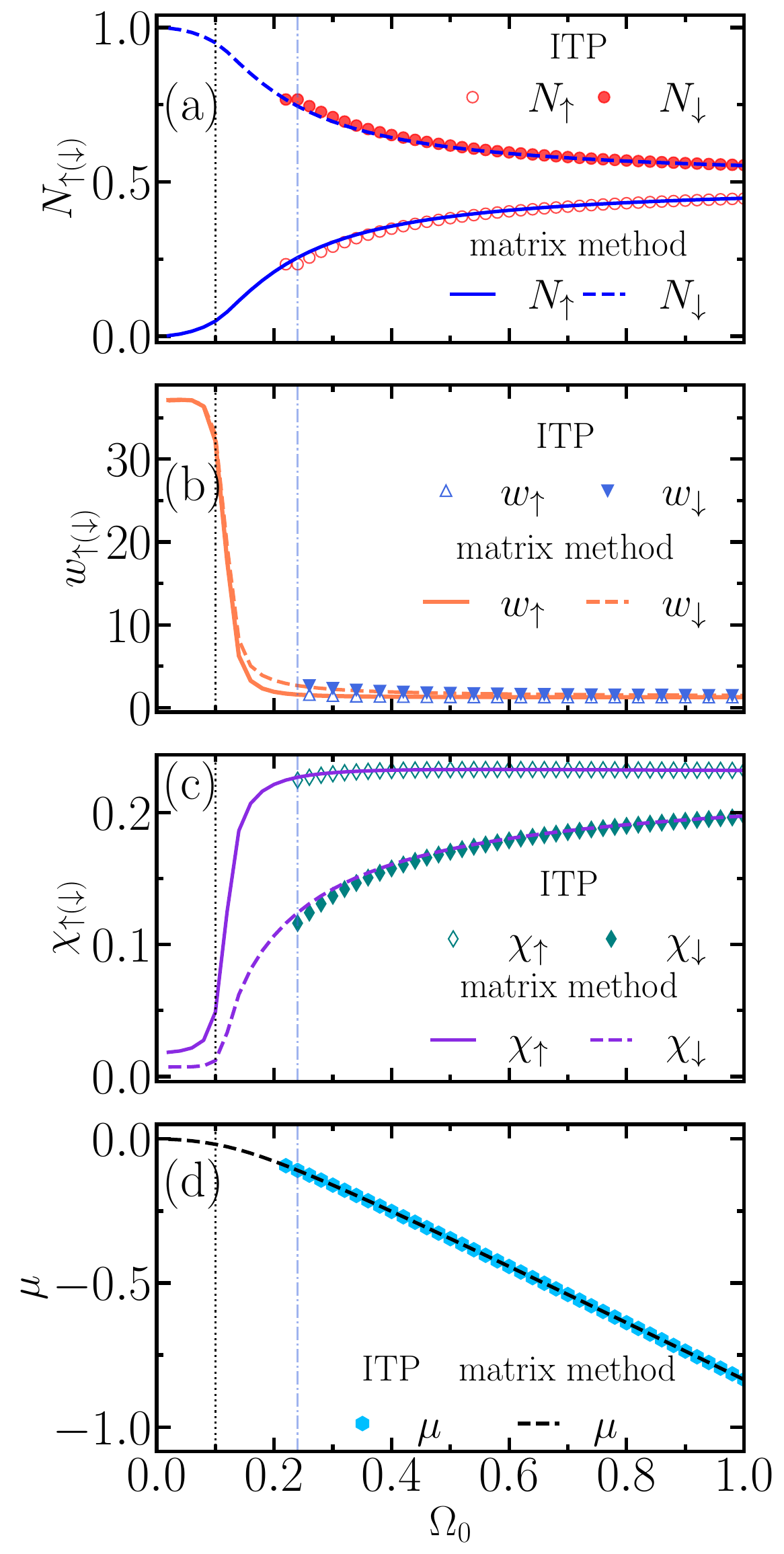}
    \caption{(a) Variation of population $N_{\upar (\dar)}$, (b) width $w_{\upar(\dar)}$, (c) IPR $\chi_{\upar(\dar)}$ and (d) chemical potential $\mu$ as a function of $\Omega_0$ at $V_2/V_1 = 0.5, V_1 = 1.0, g = 0$. Here, the solid and dashed lines represent the entities obtained with the matrix method. The quantities obtained by the ITP method are shown by markers. Increase in $\Omega_0$ results in transfer of probability from $\dar$ to $\upar$ component. The increasing IPR indicates that the spin-down and  spin-up components are tending to localize with the same profile due to the Rabi coupling, while spin-up is originally localized by the $V_{\upar}(x)$ potential.}
    \label{fig:N-w-var-omega}
\end{figure}

As the spin-up component interacts with the quasiperiodic potential, the other one is expected to be correlated with it due to the Rabi coupling acting since minimization of the Rabi energy requires similarity of these densities (see Appendix~\ref{sec:appenA} for details). To quantitatively investigate the localization induced by the Rabi coupling, in Fig.~\ref{fig:density}, we present the condensate density profile $|\psi_{\upar (\dar)}|^2$ for different values of $\Omega_{0}$. At very small $\Omega_{0} = 0.08$, both components are broadly distributed over the $[-L, L]$ range, showing different patterns.
While $|\psi_{\upar}|^2$ is located mainly in the vicinities of the minima $x_{i}$ (see Eq. \eqref{eq:Vprimes}), $|\psi_{\dar}|^2$ shows a more continuous distribution with peaks near $x_{i}$ minima [see Fig.~\ref{fig:density}(a)]. The distinct peaks of $|\psi_{\dar} (x_i)|^2$ demonstrate the effect of Rabi-coupling to couple the components even at very low values of $\Omega_0$. Conversely, for larger values of $\Omega_{0}$, both components are localized near the minima at $x=0$ [see figure \ref{fig:density}(b,c)]. With the further increase of $\Omega_{0}$,  $|\psi_{\dar}|^2$ closely follows the $|\psi_{\upar}|^2$.  
 In this context, it is instructive to compare the insets in Fig.~\ref{fig:density}(b) and Fig.~\ref{fig:density}(c, d). The inset in Fig. (c) depicts that for a moderate Rabi coupling, the BEC shows two distinct types of localization: the exponential one for the spin-down and the Gaussian one for the spin-up states, respectively. For a strong Rabi coupling, where $\Omega_0 \sim \lambda$,  both components show the Gaussian-like localization. This is in agreement with the findings of Santos and Cardoso in Ref.~\cite{Santos:2021}.

The effect of the Rabi coupling can be further understood by analyzing the population $N_{j}$ (Fig.~\ref{fig:N-w-var-omega}(a)), width $w_{j}$ (Fig.~\ref{fig:N-w-var-omega}(b)), and IPR $\chi_{j}$ (Fig.~\ref{fig:N-w-var-omega}(c)), and chemical potential $\mu$ (Fig.~\ref{fig:N-w-var-omega}(d)) of the condensate. This behavior can be compared with induced localization for the harmonic trap with the frequency $\omega_{0}$ (see Eq. \eqref{eq:Vprimes} and Appendix \ref{sec:appenA}). 

{The obtained population imbalance forms the BEC spin polarization. In spin–orbit coupled multicomponent condensates such modifications typically arise from unequal interactions lifting Manakov’s symmetry of the spin states~\cite{Sarkar:2025}. In the present case, the spin-asymmetric potential similarly induces population imbalance, so tuning the Rabi coupling offers an additional means of controlling spin states in BECs, beyond the conventional approach of tuning atomic interactions. An example of this effect is provided in Appendix~\ref{appen:C}. }

\begin{figure}[!htp]
    \centering
    \includegraphics[width=0.95\linewidth]{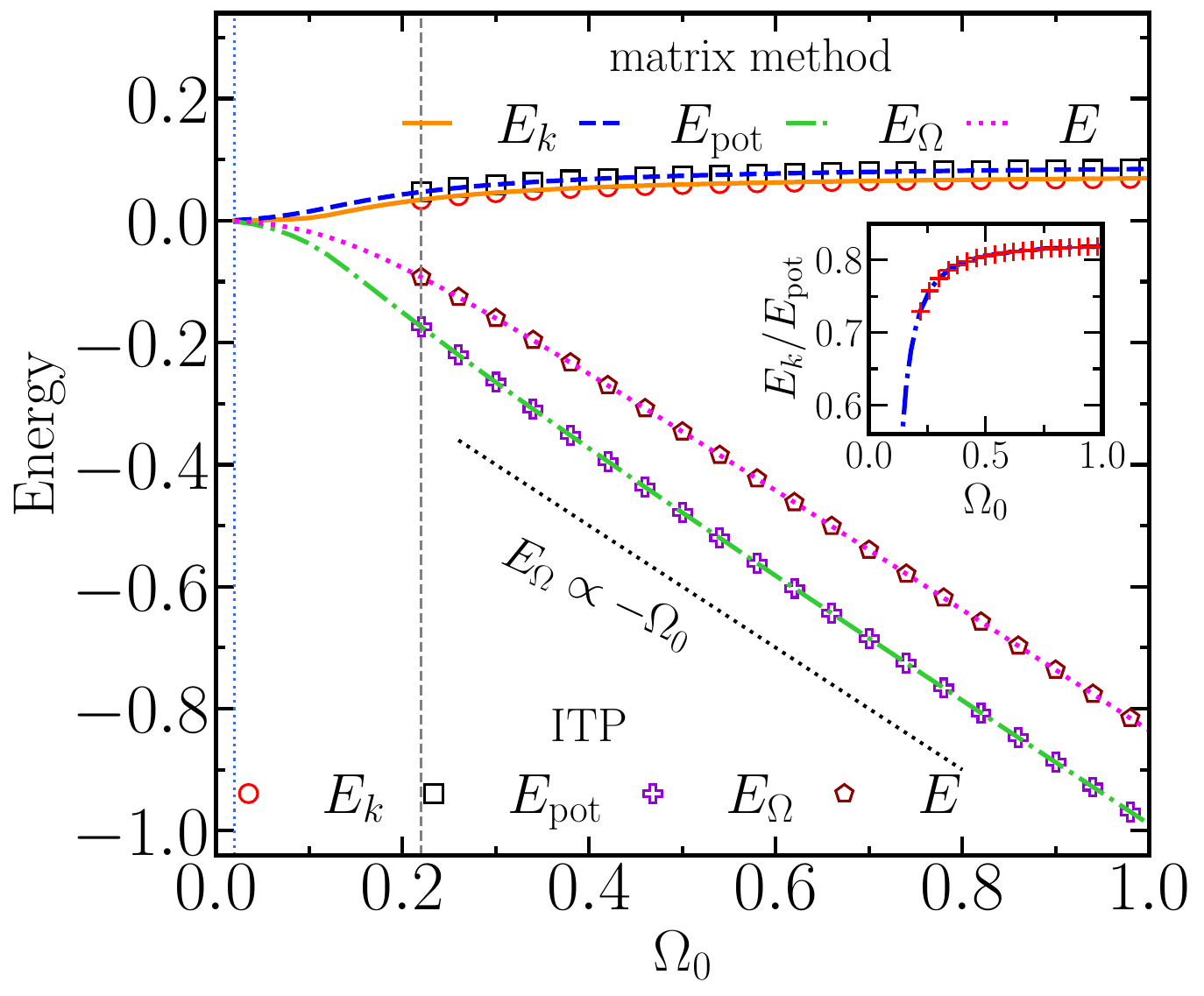}
    \caption{{Variation of different energies defined by Eqs.\eqref{eq:Epot}-\eqref{eq:EOmega} as a function of Rabi coupling $\Omega_0$ at $V_2/V_1 = 0.5, V_1 = 1.0.$ As $\Omega_0$ increases, the ratio of potential $E_{\rm pot}$ and kinetic $E_{k}$ energies approximately saturates toward $\sim 0.85,$ as shown in the inset, while the dominating negative  $E_{\Omega}$ minimizes the total energy $E=E_{\rm pot}+E_{k}+E_{\Omega}.$ The black dotted line is presented to highlight the behavior of $E_\Omega \propto -\Omega_0.$ } }
    \label{fig:Energy-omega}
\end{figure}

Figure \ref{fig:N-w-var-omega}(a) shows that for weak Rabi coupling $\Omega_{0}\lesssim 0.2$ the trapped condensate population $N_{\upar}\ll 1,$ while $N_{\dar}$ is close to 1 because at small $\Omega_0$ one has $|\psi_{\upar}|^{2}\ll\,|\psi_{\dar}|^{2},$ corresponding to the fact that minimizing the BEC energy requires a large occupation of the broad spin ${\dar}$ state.  Increasing the Rabi coupling tends to equalize the population of both components. On the other hand, condensate width is very large for $\Omega_0 < 0.1$, beyond that the decreasing width $w_{\upar(\dar)}$ reveals that the increase in $\Omega_0$ leads to mutual localization of the spin-related components eventually following each other [see Fig. \ref{fig:N-w-var-omega}(b)]. Interestingly, for the harmonic potential [see Appendix], this localization process is different because of the strong confinement of the spin-up component in the harmonic trap. Comparing width $w_{\upar (\dar)}$ and IPR $\chi_{\upar (\dar)}$ in Figs. \ref{fig:N-w-var-omega}(b,c) clearly shows that $\chi_{\dar}$ increases with $\Omega_0$, demonstrating again the Rabi-induced localization, complemented by inset semilogscale profiles in Fig.~\ref{fig:density}. In that context, the linearly decreasing chemical potential $\mu\approx -\Omega_{0}$ [Fig.~\ref{fig:N-w-var-omega}(d)] at $\Omega_{0}\geq 0.5$ defines the formation of bound state of similar spin-up and spin-down components due to a strong coupling. 
\begin{figure}[!htp]
    \centering
    \includegraphics[width=\linewidth]{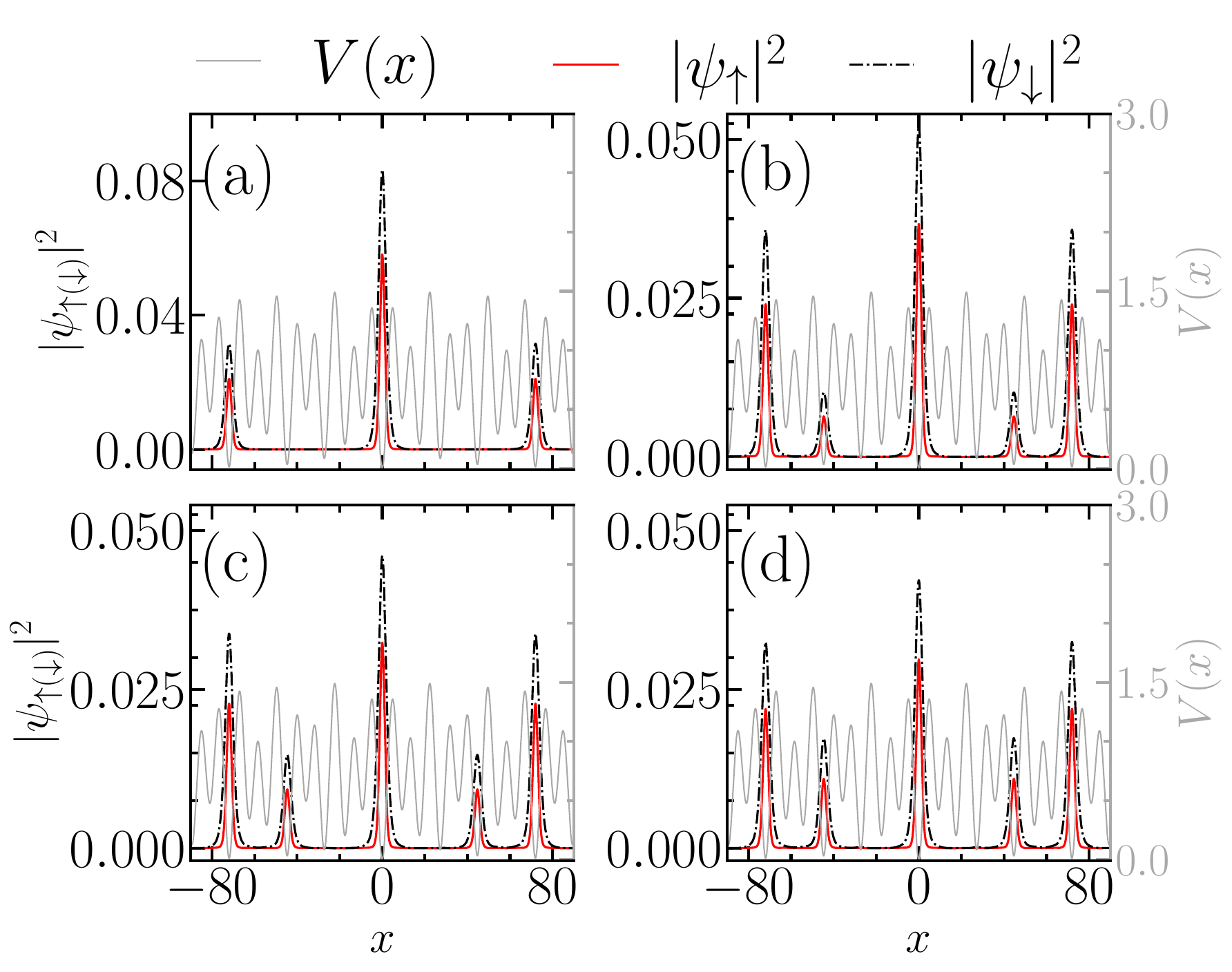}
    \caption{The condensate density $|\psi_{\upar(\dar)}|^2$ for different self-interaction strengths: (a) $g = 0.1$, (b) $g = 0.3$, (c) $g = 0.4$, and (d) $g = 0.5.$ Increasing the repulsive interaction causes the condensate to break into multiple fragments, forming several peaks at potential minima located at $x \neq 0$. The other parameters are kept as follows: $V_1 = 1.0$, $V_2 = 0.5$, $k_1 = 0.35$, $k_2/k_1 = (\sqrt{5}-1)/2$, and $\Omega_0 = 0.3.$}
    \label{fig:density-int}
\end{figure}

 In Fig.~\ref{fig:Energy-omega} we show different energies as a function of $\Omega_0$ with the other parameters the same as in Fig.~\ref{fig:N-w-var-omega}. These energies are defined as follows. The potential energy corresponding to Eq. \eqref{eq:VupVdown}:
\begin{equation}\label{eq:Epot}
E_{\rm pot} = \int_{-\infty}^{\infty} V_{\upar}(x)|\psi_{\upar}|^2 dx,   
\end{equation} 
the kinetic energy 
\begin{equation}\label{eq:Ek}
E_{k} = \frac{1}{2}\int_{-\infty}^{\infty} |\psi_{\upar}^{\prime}|^2 dx 
+ \frac{1}{2}\int_{-\infty}^{\infty} |\psi_{\dar}^{\prime}|^2 dx,
\end{equation}
and the Rabi coupling energy 
\begin{equation}\label{eq:EOmega}
E_{\Omega} = \Omega_{0} \int_{-\infty}^{\infty} (\psi^{*}_{\dar}\psi_{\upar} + \psi^{*}_{\upar}\psi_{\dar}) dx.   
\end{equation}

{For small $\Omega_{0},$ the potential and kinetic energies increase and then approximately saturate. Throughout the variation these energies are relatively close to each other. As $\Omega_0$ increases, the ratio of $E_{k}/E_{\rm pot}$ saturates toward $0.85$ as shown in the inset of Fig.~\ref{fig:Energy-omega}, while the Rabi coupling contribution $E_{\Omega},$ decreasing approximately as $-\Omega_{0},$ dominates in the total energy. The strong Rabi coupling requires $\psi_{\upar}\approx -\psi_{\dar}$ implying that in this limit localization in potential in Eq. \eqref{eq:VupVdown} is similar to localization in spin-independent potential $V_{\upar}(x) =  V_{\dar}(x) = V(x)$ (see Appendix \ref{sec:appenA} for details).}

% For small $\Omega_{0}$ the potential energy exceeds the kinetic term. As $\Omega_0$ increases, these energies become close ($E_{k}\approx E_{\rm pot}$) since the dominance of the Rabi coupling $E_{\Omega}\approx -\Omega_{0},$ decreasing approximately linearly with $\Omega_0,$ requires that $\psi_{\upar}\approx -\psi_{\dar}$ (see Appendix for details). 

%%%%%%%%%%%%%%%%%%%%%%%%%%%%%%%%%%%%%%%%%%%%%%%%%%%

\subsection{Ground state of self-interacting condensate}
\label{ground-stat-int}
In this subsection we explore the effect of the self-interaction {with lifted Manakov's symmetry described by a single parameter $g$, where  $g=g_{\upar\upar} = g_{\dar\dar}$ and $g_{\upar\dar} = g_{\dar\upar}=0,$ without nonlinear coupling between different spin components, on the ground state of the condensate.} The resulting self-interaction energy is given by:
\begin{equation}\label{eq:Eint}
E_{\rm int} = \frac{g}{2} \sum_{j = \upar,\dar} \int_{-\infty}^{\infty} |\psi_{j}|^4 dx 
=\frac{g}{2}\sum_{j = \upar,\dar}N_{j}^{2}\chi_{j}.    
\end{equation}

In Fig.~\ref{fig:density-int}, we show the condensate density for different $g$ by keeping Rabi coupling at $\Omega_{0} = 0.3.$ At $g=0$, the condensate is perfectly localized near $x = 0$ as seen in Fig.~\ref{fig:density}(b). The repulsive intra-species interactions result in expanding the condensate from the central minimum and leads to its fragmentation at various positions. For example, in Fig. \ref{fig:density-int}(a) $g = 0.1$, the BEC localized at $x = 0$ fragments with two additional peaks at $x \sim \pm 80$, where another minimum of $V(x)$ is located. With the further increase in $g$, the condensate breaks into more fragments as in Fig. \ref{fig:density-int}(c-d) fragmentation occurs with five peaks situated around $x \sim \pm 70, \pm 80,$ and $x = 0.$ 

This strong effect of self-interaction is the specific feature of the quasiperiodic potential having a variety of minima with small $V(x_{i})$ and close $\omega_{i}.$  Thus, a relatively weak self-repulsion can effectively redistribute the condensate density between these minima with similar energies. This effect of self-repulsion is enhanced by the fact that spread of spin-down component is not influenced by the quasiperiodic potential. At a moderate or strong $\Omega_{0},$ where $|\psi_{\dar}|\approx |\psi_{\upar}|,$ the critical self-interaction $g_{\rm cr}$ that begins the  occupation of the wing $x_{i}\neq 0$ minima, can be estimated as $g_{\rm cr}\chi_{0}/2\approx\,E_{i}-E_{0},$ where $\chi_{0}=\sqrt{\omega_{0}/2\pi}$ is the IPR of the state localized near $x=0$ and %$E_{i}=V(x_{i})+\omega_{i}$ 
$E_{i}$ is the closest to $E_{0}\approx\omega_{0}$ energy. Thus, even a relatively small $g\sim 0.1$ can cause density redistribution between the distant minima seen as the BEC fragmentation while the effect on $g$ on the states near the $x=0$ minimum becomes considerable at $g\sim\sqrt{2\pi\omega_{0}}\gtrsim 1.$

After analyzing the role of Rabi coupling in the induced localization of the non-interacting and interacting binary condensate now we proceed to explore the dynamics of the localized states.

\section{Dynamics of induced localized condensates}
\label{sec:4}
In this Section, we proceed to study the BEC dynamics caused by periodical driving with the time modulated Rabi frequency as:
\begin{align}
    \Omega(t) =
    \begin{cases}
        \Omega_0, & t < 0 \\  
        \Omega_0 + \Omega_1 \sin(\omega_{\rm osc} t), & t > 0
    \end{cases}
    \label{eqn12}
\end{align}
where at $t < 0$ the condensate is in the ground state.  For capturing the dynamics we use different entities such as the density, miscibility (\ref{eq:misc}), and correlation function (\ref{eqn:c_t}). To make the time modulation a (possibly strong) perturbation, we always maintain $\Omega_1/\Omega_0=1/2$ ratio and the oscillation frequency $\omega_{\rm osc}$ is varied within the interval from 0.1 to 1.0. 

\subsection{Effect of the oscillating Rabi frequency on the linear condensate}

\begin{figure*}[!htp]
\centering
 \includegraphics[width=\linewidth]{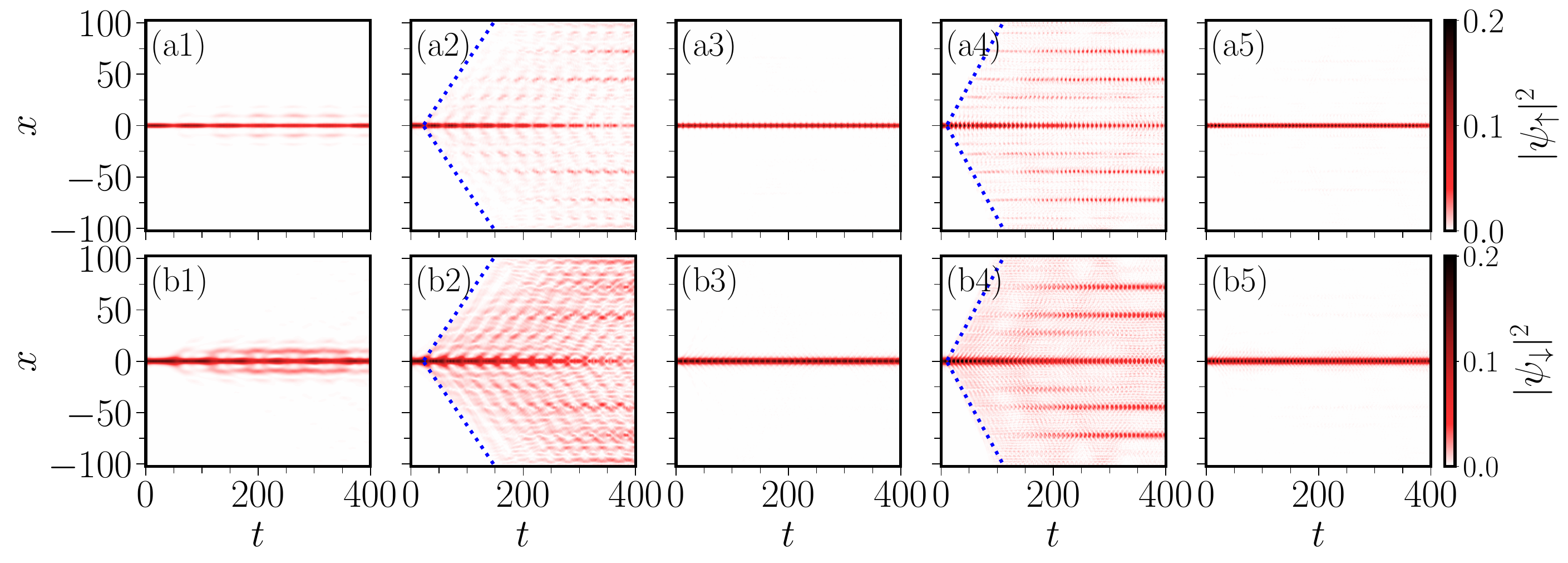}
\caption{Pseudo-colormap representation of the condensate densities $|\psi_{\upar}|^2$ (upper panel) and $|\psi_{\dar}|^2$ (lower panel) in the $(t,x)$ plane for different frequencies $\omega_{\rm osc}$: (a1, b1) $\omega_{\rm osc} = 0.1$, (a2, b2) $\omega_{\rm osc} = 0.2$, (a3, b3) $\omega_{\rm osc} = 0.4$, (a4, b4) $\omega_{\rm osc} = 0.8$, and (a5, b5) $\omega_{\rm osc} = 1.0$. For $\omega_{\rm osc} = 0.2$ (a2, b2) and $\omega_{\rm osc} = 0.8$ (a4, b4), the density expands. However, the expansion dynamics for $\omega_{\rm osc} = 0.8$ differ from those at $\omega_{\rm osc} = 0.2.$ Specifically, at $\omega_{\rm osc} = 0.8$, the density $|\psi_{\upar(\dar)}|^2$ eventually settles at different potential minima (a4, b4), whereas for $\omega_{\rm osc} = 0.2$ (a2, b2), it expands by emitting jets with a nearly uniform velocity. For other $\omega_{\rm osc}$, the condensate remained localized around $x = 0.$ The other parameters are $\Omega_0 = 0.4$, $\Omega_1 = 0.2$, and $g = 0.$ {Panels (a1) and (b1) correspond to double (triple)- minima evolution, panels (a2), (b2) and (a4),(b4) show tree-like patterns while panels (a3),(b3) and (a5),(b5) correspond to a "frozen" behavior.}}
\label{fig:t-x-density-non-int}
\end{figure*}
%%%%%%%%%%%%%%%%%%%%%%%%%%%%%%%%%%%%%%%%%%%%%%%%%%%%%%%%%%%%%%%%%%%%%%%
 \begin{figure}
    \centering
    \includegraphics[width=0.85\linewidth]{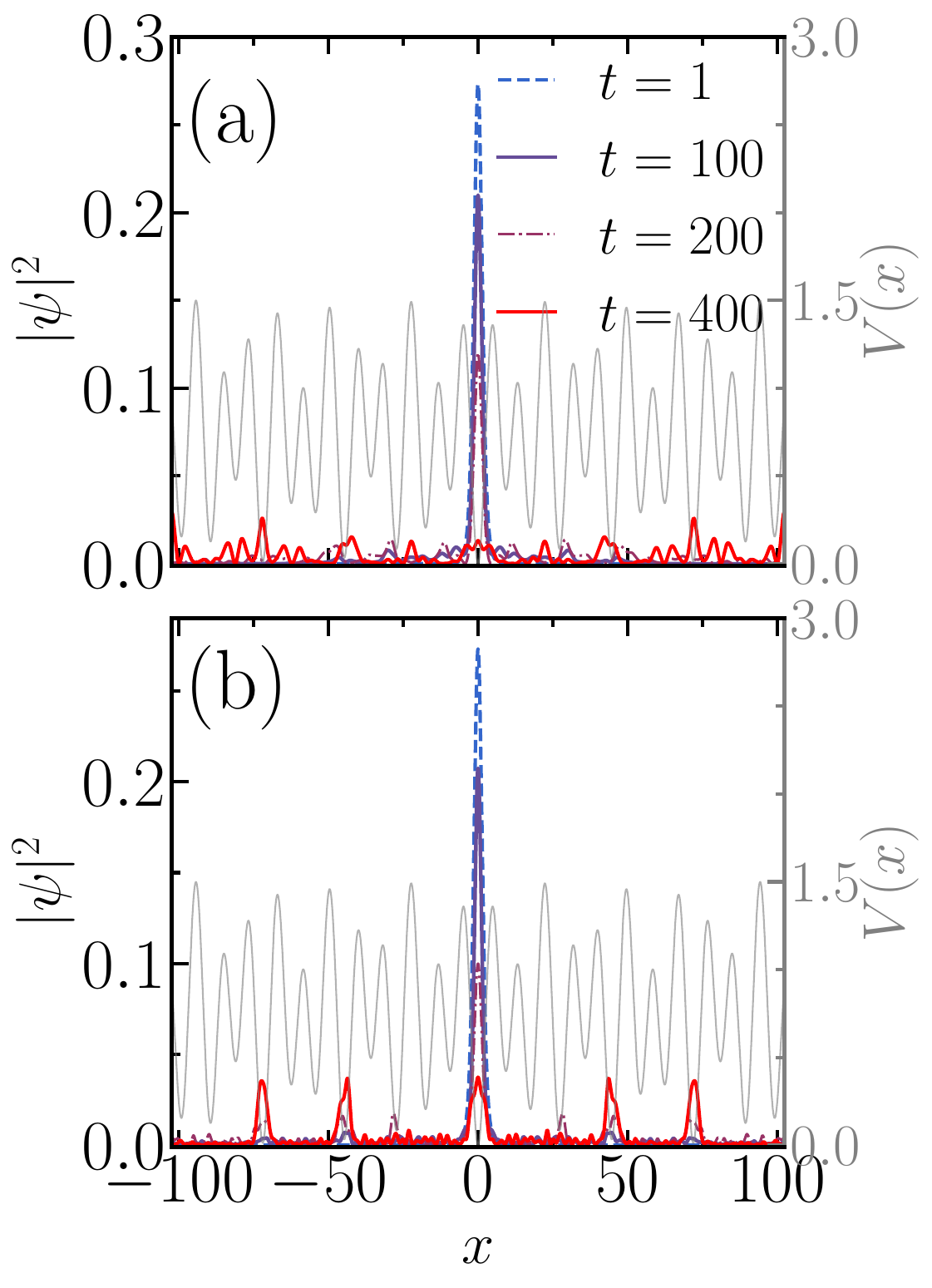}
    \caption{The total density profile $|\psi|^2 = |\psi_{\upar}|^2 + |\psi_{\dar}|^2$ is shown at different time snapshots $t = (1, 100, 200, 400)$ for (a) $\omega_{\rm osc} = 0.2$ and (b) $\omega_{\rm osc} = 0.8$. For $\omega_{\rm osc} = 0.8$, the density fragments into four distinct minima of the potential located at non-zero positions $x \neq 0$ at large time snap ($t = 400$), whereas for $\omega_{\rm osc} = 0.2$, the condensate expands.}
    \label{fig:density-t-snaps}
\end{figure}

To begin our analysis for the linear condensate, in Fig.~\ref{fig:t-x-density-non-int}, we show the evolution of the density $|\psi_{\upar(\dar)}|^2$ in the $(t,x)$ plane for oscillation frequencies ranging between $\omega_{\rm osc} = [0.1 - 1.0]$ while keeping  $\Omega_0 = 0.4$, and $\Omega_1 = 0.2$ such that $\Omega(t)$ oscillates between a relatively weak (0.2) and a relatively strong (0.6) values (cf. Fig. \ref{fig:density}). For $\omega_{\rm osc} = 0.1$ in (a1, b1), the density mainly remains localized near $x = 0,$ although, due to the potential-free spin-down component, it exhibits oscillations within two nearest minima $\pm x_{1}.$ This behavior demonstrates that the BEC acquires a relatively small energy, leading to confined oscillations within a {\it double (triple) well}.

Next, the density propagation at $\omega_{\rm osc} = 0.2$ [in (a2,b2)] exhibits driven expansion all over the space, indicating delocalization of the condensate. 
However, the effect is more pronounced for $|\psi_{\dar}|^2$ (in b2) than for $|\psi_{\upar}|^2$ (in a2) due to the confinement of the latter. The spin-down component is emitted from the central minimum  as separate jets with the velocity of the order of $\sqrt{\omega_{0}},$ corresponding to the spread a Gaussian wavepacket with the initial energy $\sim \omega_{0}$ (see Eq. \eqref{eq:Vprimes}). The decreasing of $|\psi_{\dar}|^2$ in the vicinity of the $x=0$ minimum pulls the $|\psi_{\upar}|^2$ out of this region and leads to its time-dependence and delocalization. Simultaneously, increasing in $\Omega(t)$ to $\approx \Omega_{0}+\Omega_{1}$ periodically pumps the probability from ${\upar}$ to ${\dar}$ component while lowering $\Omega(t)$ to $\approx \Omega_{0}-\Omega_{1}$ during the same period causes emission of spin-down jets and, correspondingly, a decrease in the probability for the ${\dar}-$component to be occupied.  With the gradual reduction in the densities of both components near $x=0$ point, accompanied by a decrease in their Rabi coupling, the efficiency of this process decreases. 

Similarly, for $\omega_{\rm osc} = 0.8$ [see (a4, b4)], the condensate initially expands up to $t \lesssim 200$ and starts breaking symmetrically into multiple fragments at different $x\neq\,0$ minima of $V(x).$ This feature is more clearly visible with different time snapshots of total density $|\psi|^2$ shown in Fig.~\ref{fig:density-t-snaps}. Comparison of Figs. ~\ref{fig:density-t-snaps}(a) and \ref{fig:density-t-snaps}(b) illustrates that at $\omega_{\rm osc} = 0.8$ (in (b)) the density symmetrically breaks into multiple fragments beyond $t > 200$ at $V(x)$  minima around $x \sim \pm 50, \pm 80$ and remains {\it frozen} there. In contrast, the density at $\omega_{\rm osc} = 0.2$ (in \ref{fig:density-t-snaps}(a)) expands with uniform velocity being distributed uniformly all over the space, which we define as {\it tree-} like expansion resulting in delocalization. A possible explanation for the two distinct types of delocalization phenomena is as follows: for lower oscillation frequencies such as $\omega_{\rm osc} = 0.2$, the timescale over which the condensate density expands is comparable to the semi-adiabatic timescale associated with the modulation frequency of $\Omega(t)$. This leads to a relatively modest expansion of the condensate. In contrast, at $\omega_{\rm osc} = 0.8$, the rapid variation of $\Omega(t)$ causes the condensate expansion timescale to become shorter relative to the modulation timescale of $\Omega(t)$, resulting in the condensate becoming effectively frozen at different minima. However, for other frequencies such as $\omega_{\rm osc} = 0.4$ (a3, b3) and $1.0$ (a5, b5), the condensate remains localized near $x = 0.$ At this juncture, it is worth noting that Nakamura {\it et al.}~\cite{Nakamura:2007} theoretically reported the resonant driven levitation of binary condensates subjected to two distinct harmonic traps, similarly to the Franck-Condon effect in molecular physics.

\begin{figure}
    \centering
   \includegraphics[width=\linewidth]{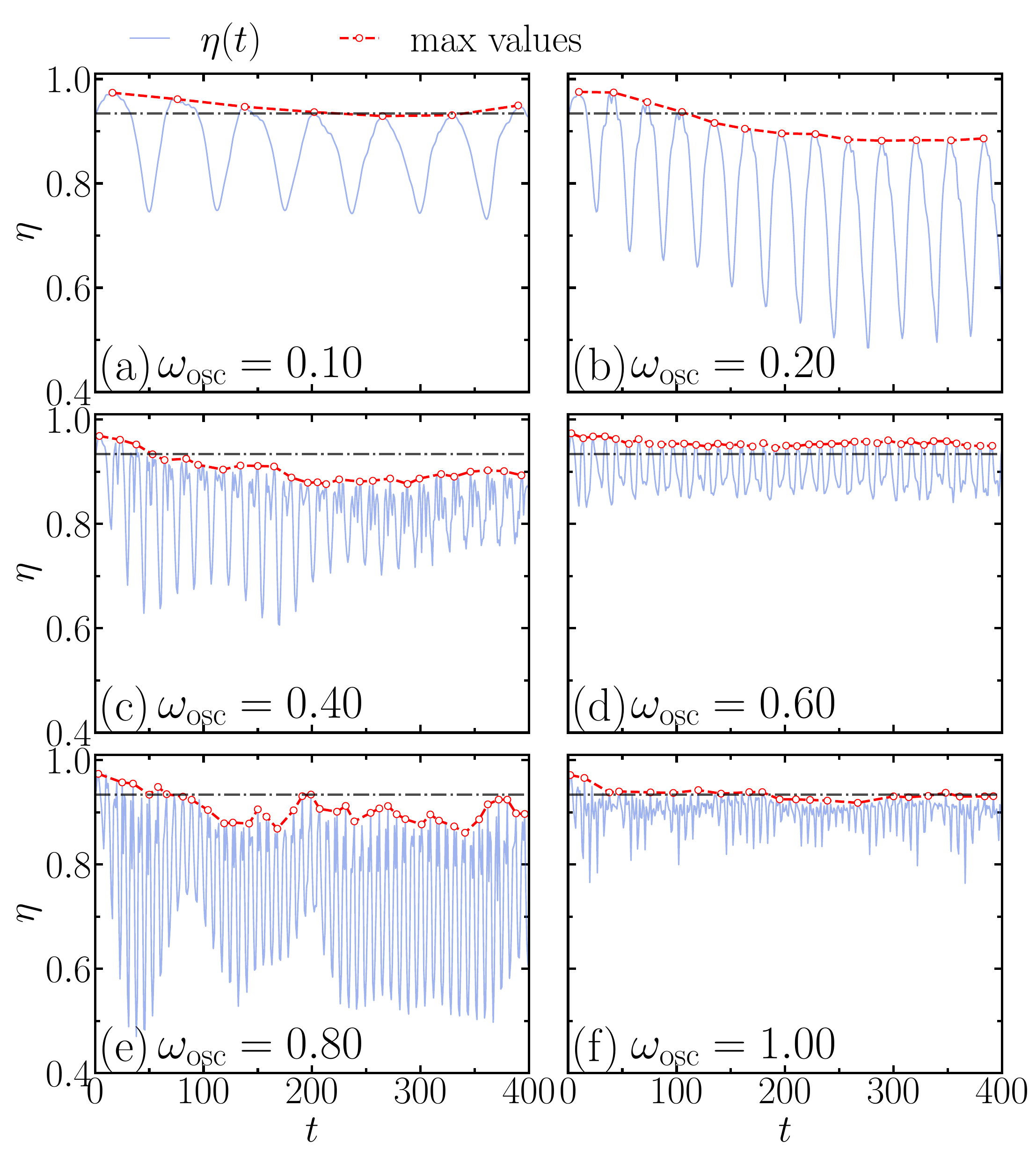}
    \caption{Time dependence of miscibility $\eta$ for different oscillation frequencies of the Rabi field $\omega_{\rm osc}$: (a) $\omega_{\rm osc} = 0.1$, (b) $\omega_{\rm osc} = 0.2$, (c) $\omega_{\rm osc} = 0.4$, (d) $\omega_{\rm osc} = 0.6$, (e) $\omega_{\rm osc} = 0.8$, and (f) $\omega_{\rm osc} = 1.0$. The black dash-dotted line indicates the initial miscibility $\eta(0) = 0.934.$ The function $\eta(t)$ represents the miscibility after turning on the Rabi field ($\Omega_1 \neq 0$). A red circle-marked dashed line highlights the maximum values of $\eta(t)$. Comparing $\eta(0)$ with the red-circled maxima shows that the miscibility gradually decreases over time for $\omega_{\rm osc} = 0.2$, $0.6$, and $0.8$, indicating expansion of the condensate.}
    \label{fig:eta-non-int}
\end{figure}
\begin{figure}[!htp]
    \centering
    \includegraphics[width=\linewidth]{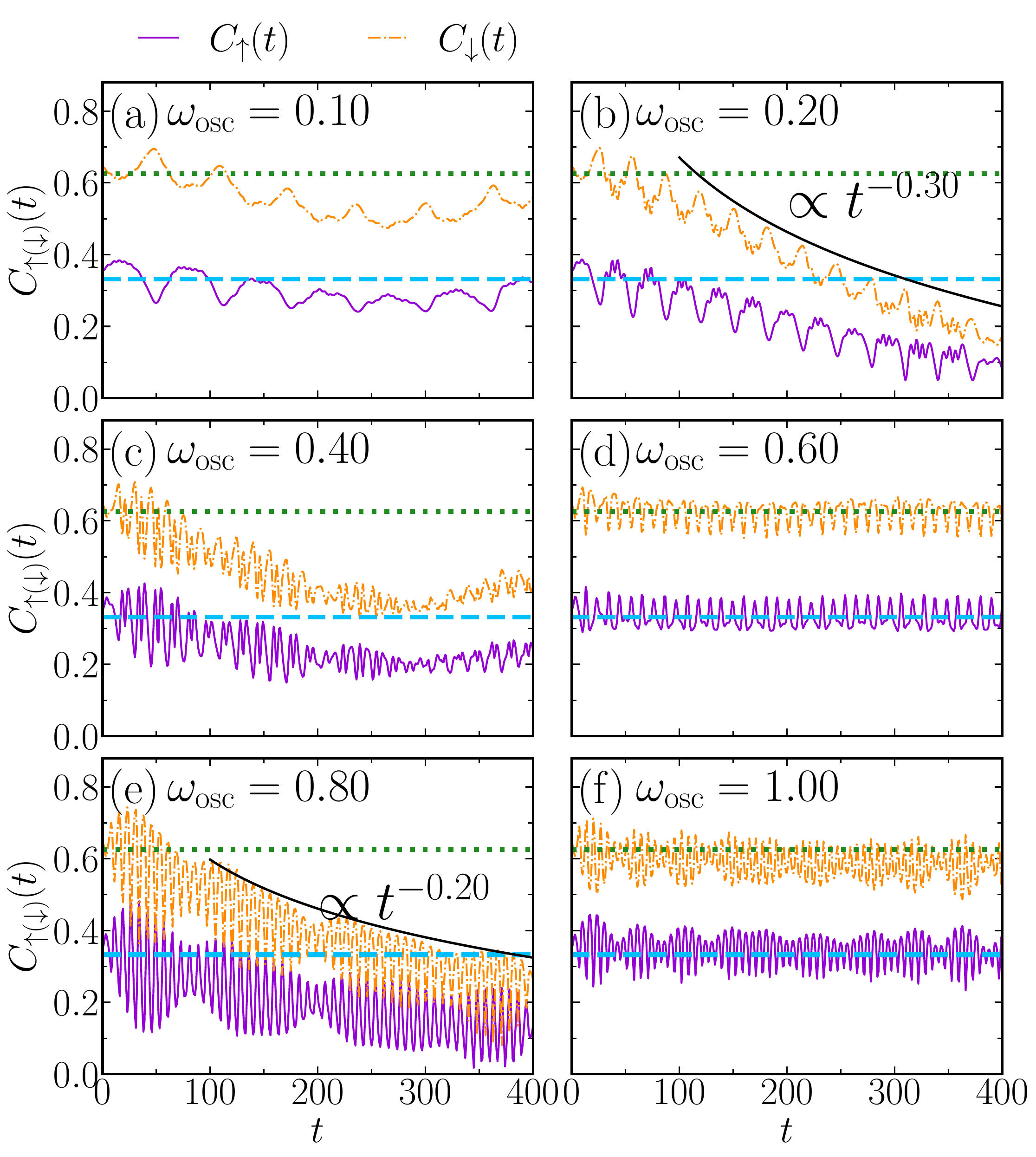}
    \caption{Temporal variation of the correlation function $C_{\upar (\dar)}(t)$ is shown for different oscillation frequencies $\omega_{\rm osc}$: (a) $\omega_{\rm osc} = 0.1$, (b) $\omega_{\rm osc} = 0.2$, (c) $\omega_{\rm osc} = 0.4$, (d) $\omega_{\rm osc} = 0.6$, (e) $\omega_{\rm osc} = 0.8$, and (f) $\omega_{\rm osc} = 1.0$. The spatiotemporal expansion of the condensate can be characterized by a power-law behavior of the form $t^{-\alpha}$ for $\omega_{\rm osc} = 0.2$ and $\omega_{\rm osc} = 0.8$, where $C_{\upar (\dar)}(t)$ exhibits a decreasing trend over time. The values of the exponent $\alpha$ are found to be $0.3$ for $\omega_{\rm osc} = 0.2$, and $0.2$ for $\omega_{\rm osc} = 0.8$. The dashed line (skyblue color), and dotted line (green color) is drawn to compare with the static case $\Omega_1 = 0$. The other parameters are same as Fig.~\ref{fig:t-x-density-non-int}}
    \label{fig:corr-non-int}
\end{figure}

Characterizing the localization and delocalization is a challenging task because of one of the components is potential-free and Rabi coupling  $\Omega_0 = 0.4$ is not strong enough. Therefore, the evolution of the density pattern cannot give much insight. In that context, we use other set of observables: miscibility $\eta(t)$ (Eq. \eqref{eq:misc}) and correlation functions $C_j(t)$ (Eq. \eqref{eqn:c_t}) to characterize these processes ~\cite{Brezi:2011,Doggen:2014,Sarkar:2023}.

Given the unequal populations of the spin components, we analyze miscibility $\eta(t)$ to characterize differences in driven localization or delocalization. Figure ~\ref{fig:eta-non-int} represents the miscibility $\eta(t)$ for the same oscillation frequencies as in Fig.~\ref{fig:t-x-density-non-int}. The maximum of $\eta(t)$ is highlighted with a red circle-marked dashed line in each panel from (a)-(f), and the black dash-dotted line is drawn at $\eta(0) = 0.934$ to indicate the extent of miscibility. In addition it should be noted that the expansion of the condensate gets manifested in the decreasing trend of $\eta(t)$ from $\eta(0),$ as depicted by the red markers in figures (b), (c), and (e) for $\omega_{\rm osc} = 0.2, 0.4$, and $0.8$, respectively.  For other cases, the miscibility remains nearly constant. Therefore, the transition from localization to delocalization can be characterized through the decreasing miscibility $\eta(t)<\eta(0).$     

Furthermore, we compute the spin-projected correlation functions $C_{j}(t)$  (see Fig.\ref{fig:corr-non-int}) for the same set of $\omega_{\rm osc}$ as in Fig.\ref{fig:eta-non-int}. Notably, $C_{\upar}(t)$ (purple solid line) and $C_{\dar}(t)$ (orange dash-dotted line) consistently maintain a close to $\pi$ phase difference for all the  cases due to periodic probability pumping from spin-$\upar$ to spin-$\dar$ component. For comparison, we also include the stationary correlation functions, represented by a sky-blue dashed line for $C_{\upar}(0)$ and a green dotted line for $C_{\dar}(0).$      
The decrease in $C_{j}(t)$ with time demonstrates the condensate escape from the ground state. Note that $C_{j}(t)$ are close to $C_{j}(0)$  for $\omega_{\rm osc} = 0.1, 0.6, 1.0$ as shown in panels (a), (d), and (f), respectively. Thus, with the course of time, the spin components remain close to the initial state in the vicinity of $x = 0,$ as mentioned earlier in Fig.~\ref{fig:t-x-density-non-int}. Conversely, for $\omega_{\rm osc} = 0.2$ [in figure (b)] and $\omega_{\rm osc} = 0.8$ [in figure (e)], the $C_{j}(t)$ decrease by following power laws with approximate exponents $t^{-0.5}$ and $t^{-0.3}$, respectively.  Also, at $\omega_{\rm osc} = 0.4 $ [in (c)], the $C_{j}(t)$ show a feeble decrement with time. Thus, the escape of the condensate can be characterized through the power-like decrease of the time-correlation functions. Also note that the relative decrease in $C_{\upar}(t)$ compared to $C_{\upar}(0)$ is less than that of $C_{\dar}(t),$ because the trapping potential in spin-up component tries to prevent its decrease. 

So far, our analysis reveals several effects of periodic Rabi frequency towards delocalization of the condensate in the absence self-interactions. Following this, in the next subsection we explore the effect of $\Omega(t)$ at self-repulsion $g>0.$

\subsection{Dynamics of induced delocalization in the presence of interaction ($g>0$)}

\begin{figure*}
    \centering
     \includegraphics[width=\linewidth]{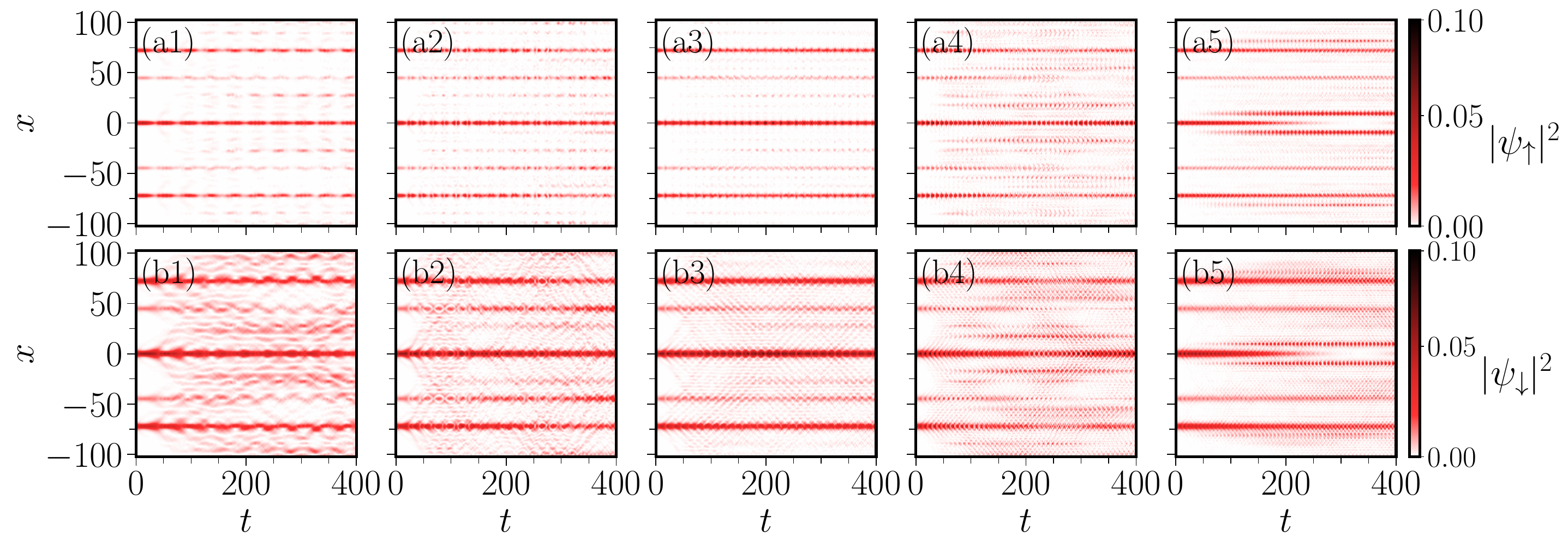}
    \caption{Evolution of the density $|\psi_{\upar}|^2$ (upper panel) and $|\psi_{\dar}|^2$ (lower panel) with the self interaction $g = 0.3$ for different $\omega_{\rm osc}$ as: (a1, b1) $\omega_{\rm osc} = 0.15$, (a2, b2) $\omega_{\rm osc} = 0.30$, (a3, b3) $\omega_{\rm osc} = 0.6$, (a4, b4) $\omega_{\rm osc} = 0.75$, and (a5, b5) $\omega_{\rm osc} = 0.9$. For different $\omega_{\rm osc}$ different types of features are visible: at $\omega_{\rm osc} = 0.15, 0.3,$ each fragmented condensate starts expanding, but at $\omega_{\rm osc} = 0.6$, fragmented parts do not show much expansion. In contrast, for other $\omega_{\rm osc} = 0.75, 0.9$, more fragments are generated with time. The other parameters are $\Omega_0 = 0.3, \Omega_1 = 0.15$. {Here panels (b1)-(b3) clearly show parquet-like patterns.}}
    \label{fig:t-x-density-g11-g22-0p3}
\end{figure*}
%%%%%%%%%%%%%%%%%%%%%%%%%%%%%%%%%%%%%%%%%%%%%%%%%%
\begin{figure}[!hb]
    \centering
    \includegraphics[width=\linewidth]{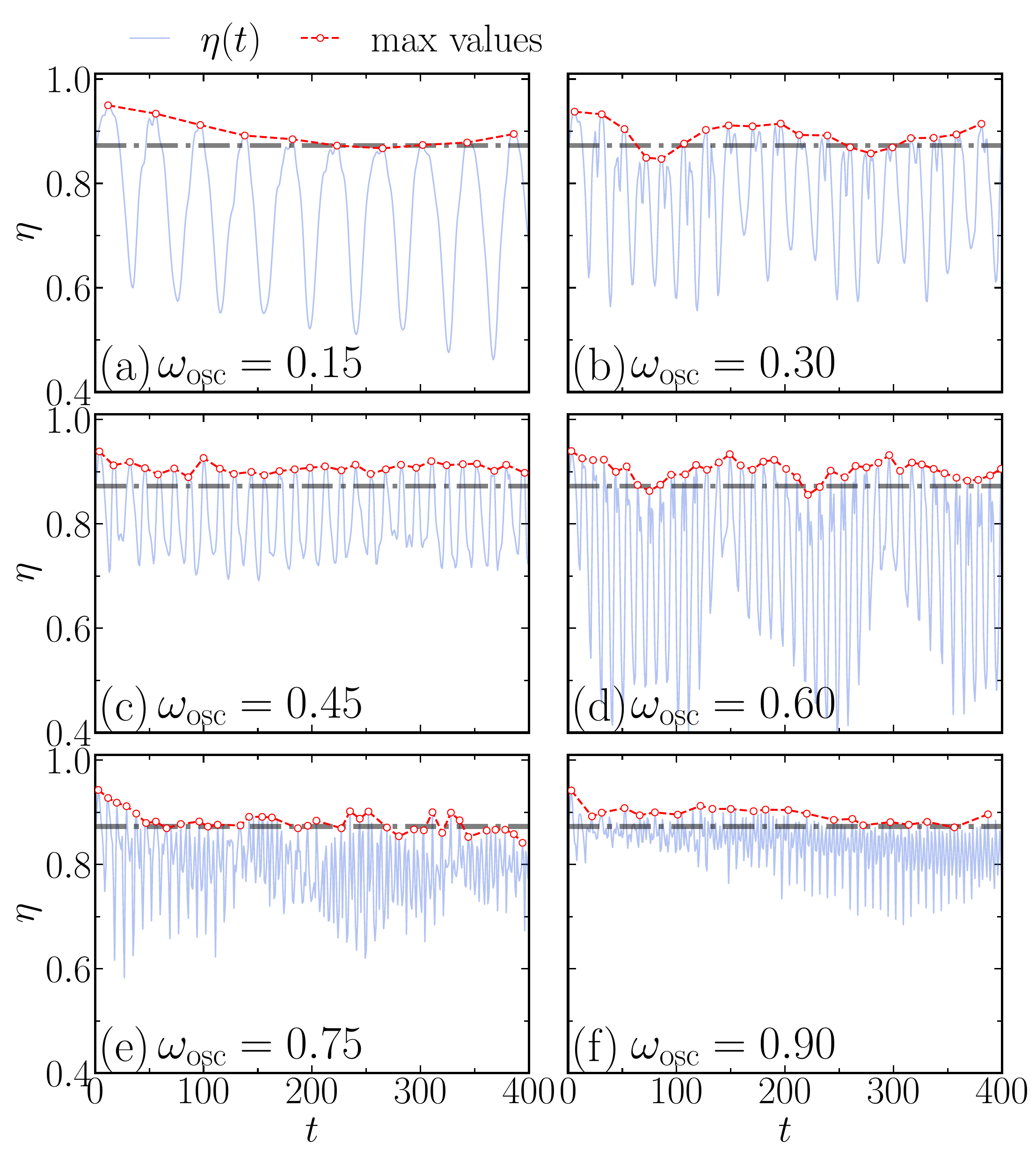}
    \caption{Evolution of miscibility $\eta(t)$ for $g = 0.3$ and different oscillation frequencies  as (a) $\omega_{\rm osc} = 0.1$, (b) $\omega_{\rm osc} = 0.2$, (c) $\omega_{\rm osc} = 0.4$, (d) $\omega_{\rm osc} = 0.6$, (e) $\omega_{\rm osc} = 0.8$, and (f) $\omega_{\rm osc} = 1.0.$. The black dash-dotted line indicates the initial miscibility $\eta(0) = 0.873$, obtained from the ground state before the time-dependent Rabi field is switched on ($\Omega_1 = 0$)
    The red circle dots show the evolution of the maximum values of $\eta(t).$}
    \label{fig:eta-t-g11-g22-0p3}
\end{figure}
%%%%%%%%%%%%%%%%%%%%%%%%%%%%%%%%%%%%%%%%%%%%%%%%%%%
As we have discussed earlier in subsection \ref{ground-stat-int}, the self-repulsions lead to fragmentation of the condensate across different potential minima and the number of fragments increases with the strength of the self-interaction. Our aim here is to analyze the dynamics of those fragmented condensates under the influence of the periodic Rabi frequency.

To begin with, in Fig.~\ref{fig:t-x-density-g11-g22-0p3} we present the evolution of the  densities $|\psi_{\upar (\dar)}|^2$ for different $\omega_{\rm osc},$ while keeping $g = 0.3,$ $\Omega_0 = 0.3$, and $\Omega_1 = 0.15.$ Initially at $t = 0$, five distinct fragments are located around $x = {0, \pm 25, \pm 75}$ [see Fig.\ref{fig:density-int}(b)]. At $\omega_{\rm osc} = 0.15$ (a1, b1), the condensate expands by jet emission from each of these fragments. At $\omega_{\rm osc} = 0.3$ (a2, b2), the expansion becomes less pronounced compared to the previous case. In contrast, for $\omega_{\rm osc} = 0.6,$ the fragmented densities remain localized at their respective positions, exhibiting no significant expansion over time. However, for higher frequencies such as $\omega_{\rm osc} = 0.75$ and $0.9$, the condensate symmetrically breaks into more fragments as time progresses. Since interactions inherently induce the BEC fragmentation, the exact identification of dynamically localized and delocalized behavior becomes even more challenging than in the non-interacting case.

To quantify different regimes, in Fig.~\ref{fig:eta-t-g11-g22-0p3} we show the miscibility $\eta(t)$ for different $\omega_{\rm osc}.$ Unlike the non-interacting case, here the decrease in $\eta(t)$ from $\eta(0)$ does not provide significant information into the condensate expansion. Nevertheless, the amplitude of $\eta(t)$ provides a qualitative explanation of the phenomena. For instance, in Fig.\ref{fig:eta-t-g11-g22-0p3}(a), the amplitude of $\eta(t)$ lies in the range $0.5 \lesssim \eta(t) \lesssim 0.9.$ Similarly, for $\omega_{\rm osc} = 0.3$ and $0.6$ [Figs.~\ref{fig:eta-t-g11-g22-0p3}(b, d)], $\eta(t)$ oscillates between $0.6 \lesssim \eta(t) \lesssim  0.9$ and $0.4 \lesssim \eta \lesssim 0.9$, respectively. On the other hand, for higher frequencies, $\eta(t)$ remain within a narrow interval, i.e. $0.7 \lesssim \eta  \lesssim 0.9.$ However, the large amplitude variation of $\eta(t)$ demonstrates the expansion from the condensate's initial positions, whereas, the small amplitude variation signifies no as such expansion of the condensate from their respective positions.     

\begin{figure}[!ht]
    \centering
    \includegraphics[width=\linewidth]{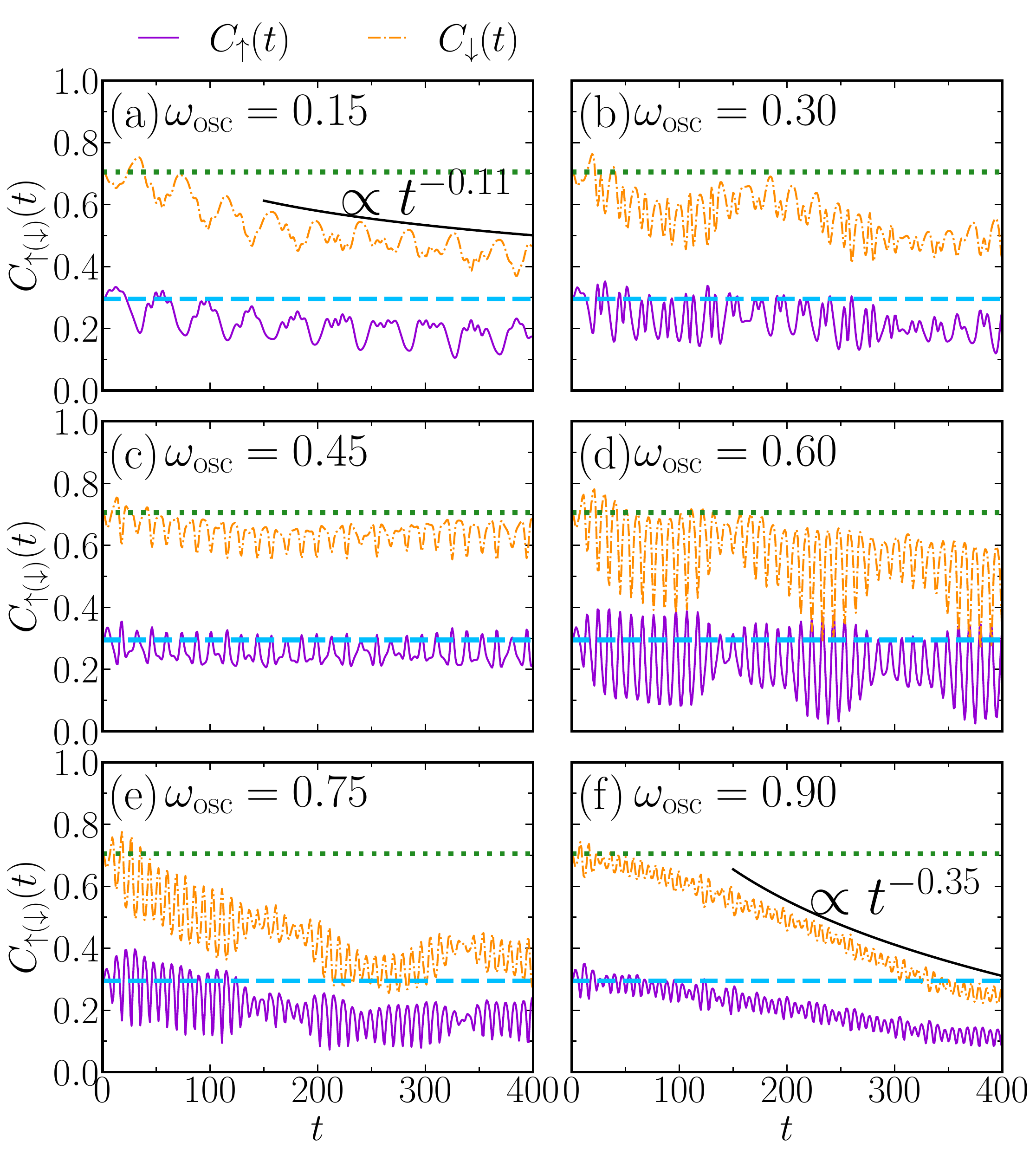}
    \caption{Time evolution of the correlation function $C_{\upar (\dar)}(t)$ is shown at $g = 0.3, \Omega_0 = 0.3, \Omega_1 = 0.15$ for different oscillation frequencies $\omega_{\rm osc}$: (a) $\omega_{\rm osc} = 0.15$, (b) $\omega_{\rm osc} = 0.3$, (c) $\omega_{\rm osc} = 0.45$, (d) $\omega_{\rm osc} = 0.6$, (e) $\omega_{\rm osc} = 0.75$, and (f) $\omega_{\rm osc} = 0.9$. Here, the dashed line (skyblue color), and dotted line (green color) is drawn to compare with the static case $\omega_{\rm osc} = 0$. The expansion of the condensate can be characterized by highlighting the power-law decrement from $C_j(0).$}     
    \label{fig:corr-g11-g22-0p3}
\end{figure}
%%%%%%%%%%%%%%%%%%%%%%%%%%%%%%%%

Furthermore, we examine the correlation function $C_j(t)$ in Fig.~\ref{fig:corr-g11-g22-0p3} for the same set of $\omega_{\rm osc}$ as shown in Fig. \ref{fig:eta-t-g11-g22-0p3}. The values of $C_j(t)$ at $\omega_{\rm osc} = 0$ are indicated by a blue dashed line at $C_{\upar} \approx 0.3$ and a green dotted line at $C_{\dar} \approx 0.7$ in each panel. Notably, at $\omega_{\rm osc} = 0.15$ (panel (a)), the power-law decay of $C_j(t)$ reflects the expansion of the density from its initial distribution. The observed power-law behavior follows an exponent of $t^{-0.11}$. Similarly, for $\omega_{\rm osc} = 0.3$ (panel (b)), although $C_j(t)$ decreases, the absence of a power-law trend indicates the suppression of condensate expansion as we have seen in the non-interacting case (Fig.~\ref{fig:corr-non-int}). Similarly, at $\omega_{\rm osc} = 0.45, 0.6$ (panels (c,d)), the decrement of $C_j(t)$ is further suppressed. On the other hand, for $\omega_{\rm osc} = 0.75, 0.9$, the relatively larger power-law decay rate demonstrates the expansion and possibly further fragmentation of the condensate.

As a result, we observe that the overall influence of the periodic Rabi frequency on the condensate is qualitatively similar to that in the non-interacting case. However, the interference of spin jets from different $\pm x_{i}$ minima makes the {\it tree-like} expansion pattern in the non-interacting case similar to a much richer {\it parquet-like} pattern in the presence of interactions.

\section{Conclusion and future outlook}
\label{sec:conclusion}

We have systematically investigated the localization and driven dynamics of a Rabi-coupled Bose-Einstein condensate subjected to a quasiperiodic potential in one spin component while the other one is potential-free. By solving the Gross-Pitaevskii equation, we have explored how the Rabi coupling, the potential, and nonlinear interactions jointly influence the ground state and the spatiotemporal characteristics of the condensate dynamics. For the linear condensates the results obtained by the imaginary time evolution align well with those obtained from the eigenmode analysis, highlighting the robustness of the induced localization mechanism.

In the absence of nonlinear interactions, we observe that the induced localization, where each component influences localization on the other, occurs when Rabi coupling exceeds a threshold value. With the introduction of self-interactions, both components exhibit localized and fragmented structures of the condensate.

After obtaining the ground state, we have studied the dynamics of the condensate under a periodically modulated Rabi frequency. When the driving frequency resonates with the intrinsic excitation modes of the localized system, we observe dynamically induced delocalization in both components accompanied by population redistribution. At higher harmonics of the resonant frequency, the condensate transitions into fragmented states, indicating mode-selective excitation. To quantitatively characterize these dynamical states, we have analyzed the evolution of miscibility and the temporal correlation functions of the spin components. Notably, similar features of drive-induced delocalization persist in the presence of interactions.

Here we propose a feasible experimental scheme to realize spin-dependent potentials in pseudospin-1/2 condensates. To generate such potentials, Bragg diffraction can be employed to selectively "tune out" specific wavelengths of optical lattices~\cite{Meng:2023}. The experiment begins by creating a  BEC trapped under a superimposed optical lattice potential $U_{1}(x)$ and $U_{2}(x).$ Initially, atoms are condensed in one of the hyperfine states, and a short laser pulse is applied, causing Bragg diffraction into higher momentum states. Subsequently, the lattice wavelengths are finely tuned so that atoms in the spin-up component experience only $U_{1}(x)$, while those in spin-down interact solely with $U_{2}(x)$. This approach has previously been used to realize binary BECs in spin-dependent twisted-bilayer lattices~\cite{Meng:2023}. For our model, once the condensate is loaded into spin-dependent lattices, one of the lattice potential is slowly ramped down to avoid excitations, allowing that component to become free from trapping while the other is confined with the optical lattice. Finally, an external magnetic field that couples the components acts as the linear Rabi coupling for the GP model.

The induced localization and drive-induced delocalization explored in this work open up several promising directions for future research. One particularly intriguing avenue involves the competitive interplay between non-commuting spin-orbit and Rabi couplings on the condensate density components. Studying their combined effects within this hybrid setup could yield rich physics. Additionally, various sets of intra- and interspecies interactions may produce phases with separated components~\cite{Sarkar:2025}, thereby introducing new dynamical behaviors under time-modulated Rabi driving. In addition to these theoretical proposals, our findings help to engineer hybrid trapping of BECs in experiments where one component can be controlled by tuning the other component when they are coupled. Also, it can be of interest to design an experiment to probe a controlled induced localization-delocalization transition in ultracold atomic gases. 

\section*{acknowledgments}
SKS would like to acknowledge the supercomputing facilities Param-Ishan and Param-Kamrupa at IITG, where all numerical simulations are performed. The work of E Y S is supported through Grants No. PGC2018-101355-B-I00 and PID2021-126273NB-I00 funded by MIUCI/AEI/10.13039/501100011033 and by the ERDF 'A way of making Europe', and by the Basque Government through Grant No. IT1470-22.

% \newpage
\appendix

\counterwithin{figure}{section}
\section{Analytical approaches and scaling analysis}
\label{sec:appenA}
\begin{figure}[!htp]
    \centering
    \includegraphics[width=0.85\linewidth]{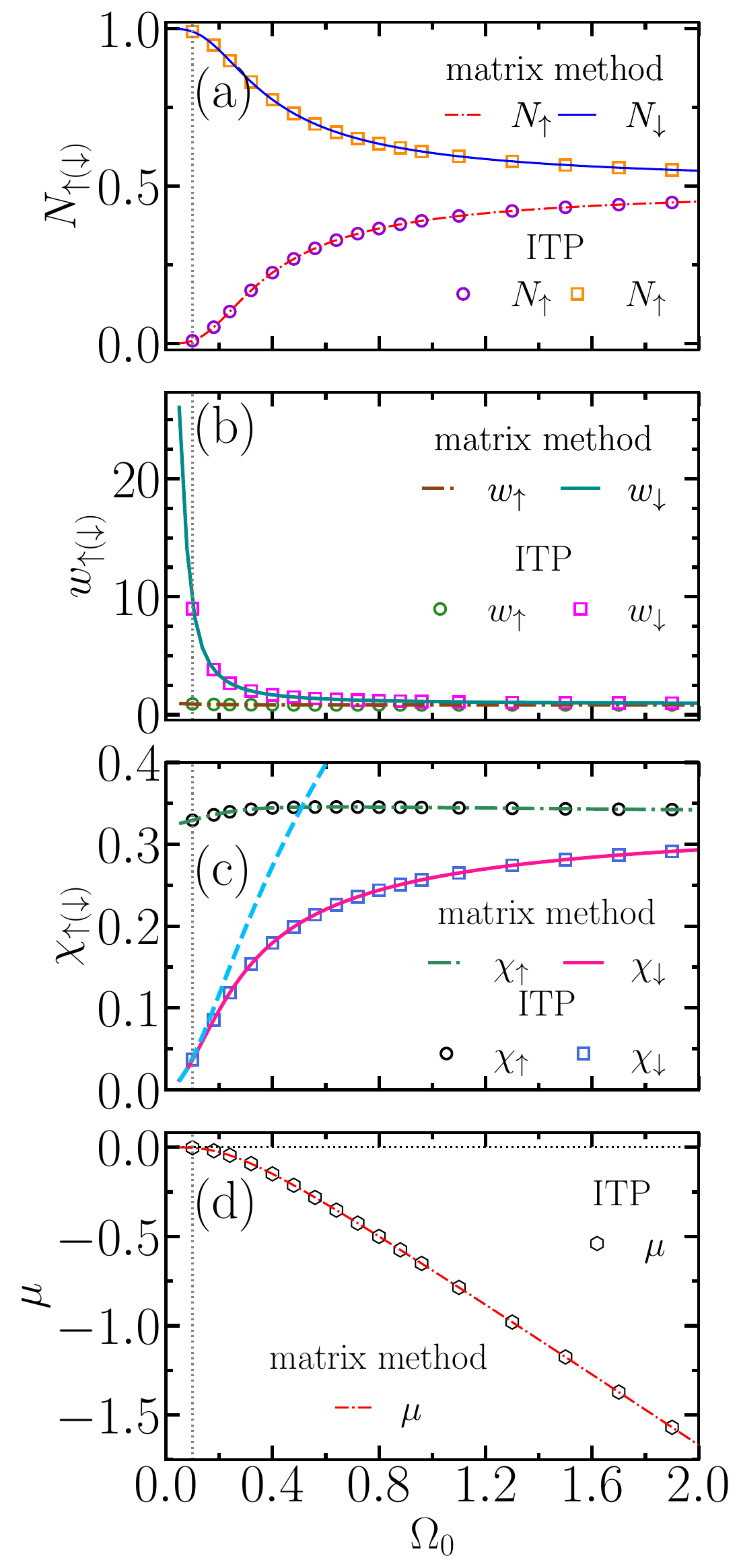}
    \caption{(a) Variation of population as a function of Rabi coupling $\Omega_{0}$ for $V=V_{\upar}(x) = x^2/2.$ (b) Width $w_{\upar(\dar)},$ (c) IPR $\chi_{\upar(\dar)},$ and (d) the chemical potential as a function of $\Omega_{0}$ at $g = 0.$ The increasing IPR indicates that the condensate of spin-down component tends to localize due to the interaction with spin-up one. In panel (c), the dashed cyan line is drawn to show the comparison with $\chi_{\dar}=\sqrt{|\mu|/2}$ corresponding to $\psi_{\dar}$ in Eq. \eqref{eqn:A2}.}
    \label{fig:N-w-var-omega-HO}
\end{figure}

Here we consider analytical approaches to the induced by the Rabi coupling localization for several realizations of interest. As in the main text, we assume self interaction $g_{\upar\upar}=g_{\dar\dar}=g$ with no cross-spin coupling and present the GPEs as: 

\begin{subequations}%
%\label{eqn1}%
\begin{align}%
% \begin{split}
 \mu\psi_{\upar} = & \left[-\frac{1}{2}\frac{d^{2}}{dx^{2}} + 
 g \vert \psi_{\upar} \vert ^{2}  
 + V_{\upar}(x)\right]\psi_{\upar} + \Omega_{0} \psi_{\dar}, \label{stat1(a)}\\ 
 \mu\psi_{\dar}  = & \left[-\frac{1}{2}\frac{d^{2}}{dx^{2}} + 
 g\vert \psi_{\dar} \vert ^{2} \right]\psi_{\dar}  + \Omega_{0} \psi_{\upar}. \label{stat1(b)} 
\end{align}
\label{eq:GP-stat}
\end{subequations}
Below we consider a harmonic trap $V_{\upar}(x)=\lambda^{2}x^{2}/2-V_{0},$ where $V_{0}$ is a uniform shift, which can influence the total wavefunction while acting on one spin component only. We present the wave function in the form
with $\psi_{\upar}=\sqrt{N_{\upar}}\phi_{\upar}(x),$ $\psi_{\dar}=\sqrt{N_{\dar}}\phi_{\dar}(x),$ where functions $\phi_{\upar}(x)$ and $\phi_{\upar}(x)$ are normalized to 1. Next, we consider how the spin-projected states behave in the case of strong $\Omega_{0}>\lambda$ and weak $\Omega_{0}\ll\,\lambda$ Rabi couplings.  General results presented in Fig. \ref{fig:N-w-var-omega-HO}  will be discussed below in terms of strong and weak couplings and briefly connected to the results for the quasiperiodic potential in the main text.   

\subsection{Strong Rabi coupling}

We begin with the very strong Rabi coupling $\Omega_{0}\gg\lambda,$ where one expects $\langle \sigma_{x}\rangle = -1$ with $N_{\upar}=N_{\dar}=1/2$ and $\phi_{\dar}(x)=-\phi_{\upar}(x).$  
We introduce the ansatz $\phi^{\rm [lim]}(x)=\phi_{\dar}(x)=-\phi_{\upar}(x)=\exp(-x^{2}/2a^{2})/\sqrt{a}\pi^{1/4}$ and minimize the total energy with respect to $a$ to obtain the ground state. The spin-diagonal kinetic energy (see Eq. \eqref{eq:Ek}) for the spin-down state described by $\phi^{\rm [lim]}(x)$ is $1/8a^{2}$ and the total (sum of the kinetic and potential energies given by Eq. \eqref{eq:Epot}) for spin-up state is $1/4a^{2}+\lambda^{2}a^{2}/8.$ By minimizing their sum we obtain $a_{0}=2^{1/4}/\sqrt{\lambda}$ and, as a result, Eqs. \eqref{eq:Epot} and \eqref{eq:Ek} yield for this state $E_{k}=E_{\rm pot}=\lambda/4\sqrt{2}.$  

Closely to this limit, we obtain corrections with respect to the small $\lambda/\Omega_{0}\ll\,1,$ where $N_{\dar}\neq N_{\upar}$ with $\phi_{\upar}(x)=\exp(-x^{2}/2a_{\upar}^{2})/\sqrt{a_{\upar}}\pi^{1/4}$ and $\phi_{\dar}(x)=\exp(-x^{2}/2a_{\dar}^{2})/\sqrt{a_{\dar}}\pi^{1/4}$ with $a_{\upar}=a_{0}.$ Thus, we obtain $N_{\dar}-N_{\upar}=\sqrt{2}\lambda/8\Omega_{0}$ and $(a_{\dar}-a_{0})/a_{0}=\lambda/2\sqrt{2}\Omega_{0},$ demonstrating that the corrections are linear in the small $\lambda/\Omega_{0}$ ratio. This asymptotic behavior matches well Fig. \ref{fig:N-w-var-omega} with $\omega_{0}\approx 0.54$ (see Eq. \eqref{eq:Vprimes}) taken as the $\lambda.$ 

Next, we briefly discuss the effect of self-repulsion $g$ on the state corresponding to $\phi^{\rm [lim]}(x)$ (see Eq. \eqref{eq:Eint}). According to this equation, here the total spin-diagonal contribution to the energy acquires the term $E_{\rm int}=g\chi^{\rm [lim]}/4,$ where $\chi^{\rm [lim]}=1/\sqrt{2\pi}a_{0}$ is the corresponding IPR. Minimization of total energy yields an increase in the width $\delta a = g/16\sqrt{\pi}\lambda$  with a relatively small effect of the nonlinearity. It increases the width of the state and, therefore, increases the spin-diagonal energy difference, decreasing the effect of a strong Rabi coupling on the state disproportion. 

\subsection{Weak Rabi coupling}

\begin{figure}[!ht]
    \centering
    \includegraphics[width=0.85\linewidth]{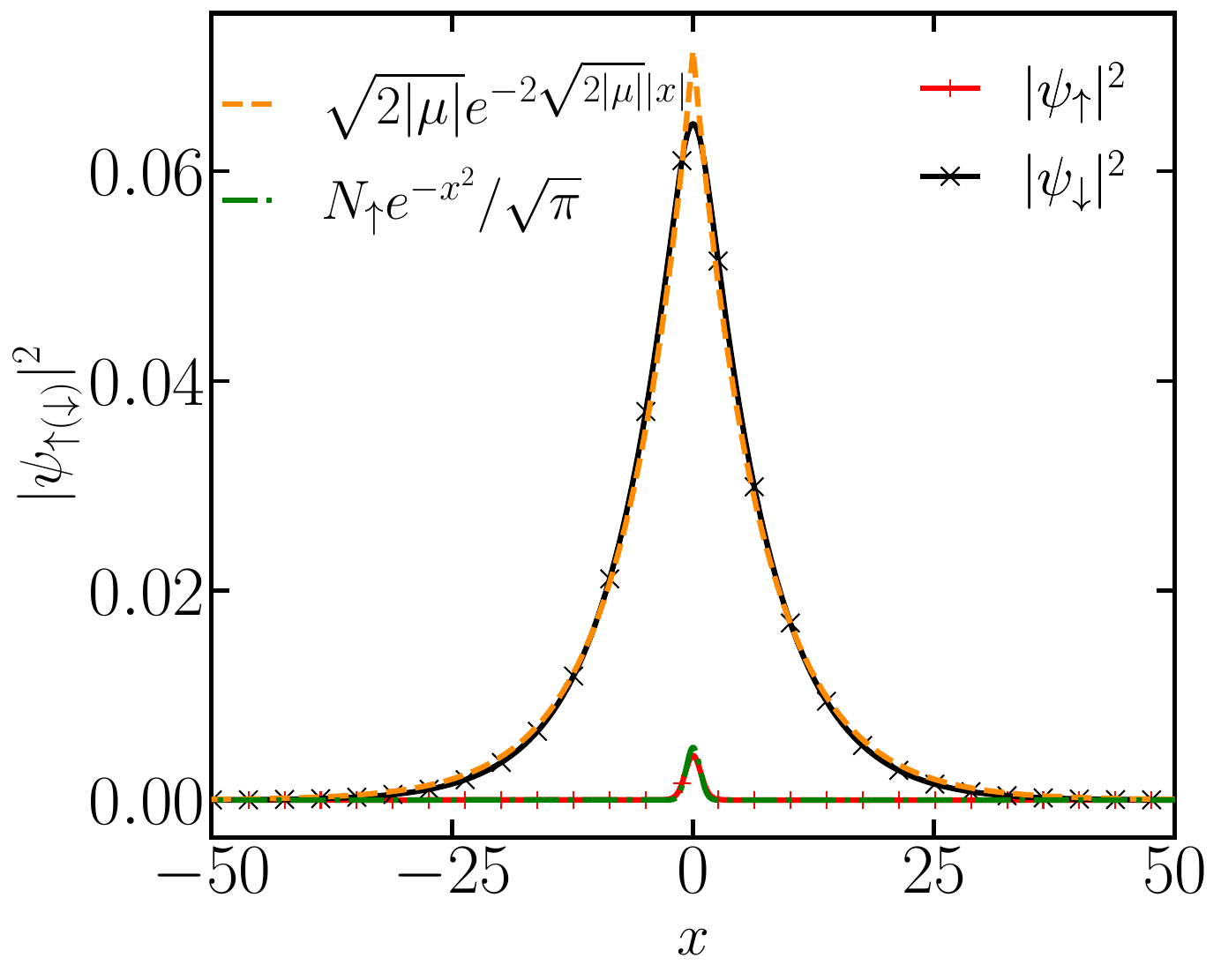}
    \caption{The condensate density profile at $\Omega_0 = 0.1$ with the harmonic trap configuration as: $V_{\upar}(x) = x^2/2, V_{\dar}(x) = 0.$ The numerical density profiles $|\psi_{\upar(\dar)}|^2$ are compared with analytical expressions for harmonic oscillator ($|\psi_{\upar}|^{2}$) and exponential localization ($|\psi_{\dar}|^{2}$).  }
    \label{fig:density-HO}
\end{figure}
Here we consider a weak Rabi coupling $\lambda/\Omega_{0}\ll 1$ with nonzero $V_{0}.$ 
We begin with assuming the spin-down wavefunction in the form
\begin{equation}
\psi_{\dar}=-2^{1/4}|\mu|^{1/4}\exp\left(-\sqrt{2|\mu|}|x|\right), 
\label{eqn:A2}
\end{equation}
normalized to $N_{\dar}=1$ with the sufficient accuracy $\sim (\Omega_{0}/\lambda)^{4},$ and the corresponding $\chi_{\dar}=\sqrt{|\mu|/2}.$ Then, the $\Omega_{0}\psi_{\upar}$ with $\psi_{\upar}=\sqrt{N_{\upar}}\phi_{0}(x),$ where 
$\phi_{0}(x)=\lambda^{1/4}\exp(-x^{2}\lambda/2)/\pi^{1/4}$ is the ground state wavefunction in the $\lambda^{2}x^{2}/2$ potential is considered as the source of a weak perturbation potential localizing $\psi_{\dar}$ on the spatial scale $1/\sqrt{|\mu|}\gg 1/\sqrt{\lambda},$  with an example shown in Fig. \ref{fig:density-HO}. Thus, we can write Eq. \eqref{stat1(b)} in the form:
\begin{equation}
\mu\psi_{\dar}  =  -\frac{1}{2}\frac{d^{2}}{dx^{2}}\psi_{\dar} + \frac{\Omega\sqrt{N_{\upar}}}{2^{1/4}|\mu|^{1/4}}\phi_{0}(x)\psi_{\dar}    
\end{equation}
and obtain for a weak narrow potential~\cite{Landau_Book}. 

\begin{equation}
\label{eq:mu1}
\mu  =  -\frac{1}{2\sqrt{2}}\frac{\Omega_{0}^{2}N_{\upar}}{|\mu|^{1/2}}
\left(\int_{-\infty}^{\infty}\phi_{0}(x) dx\right) ^{2},
\end{equation}
resulting in the relation between $\mu$ and $N_{\upar}:$ 
\begin{equation} \label{eq:mu2}
\mu  =  -\left(\sqrt{\frac{\pi}{2}}\frac{\Omega_{0}^{2}}{\lambda^{1/2}}N_{\upar}\right)^{2/3}.
\end{equation}
Next, to obtain second relation between $\mu$ and $N_{\upar},$ we use at $\sqrt{|\mu|}|x|\ll 1,$ Eq. \eqref{stat1(a)} with Eq. \eqref{eqn:A2} in the form:
\begin{equation} \label{eq:mu2}
\mu\sqrt{N_{\upar}} \phi_{0}(x) = \left(\frac{\lambda}{2}+V_{0}\right)\sqrt{N_{\upar}}\phi_{0}(x) -  2^{1/4} \Omega_{0} |\mu|^{1/4}.
\end{equation}
Multiplying both sides of Eq. \eqref{eq:mu2} by $\phi_{0}(x)$ and integrating we obtain after neglecting the left-hand side of this equation as the higher-order term in $\Omega_{0}:$ 
\begin{equation} \label{eq:Nmu}
N_{\upar} =32\pi\frac{\Omega_{0}^{4}}{(\lambda+2V_{0})\lambda^{3}};\qquad
\mu = - 8\pi\frac{\Omega_{0}^{4}}{(\lambda+2V_{0})^{2/3}\lambda^{7/3}}.
\end{equation}

Notice that at $V_{0}=0$ we obtain a simple relation: $\mu = - \Omega_{0}N_{\dar}/4$ in agreement with numerical calculations. The difference between spin-dependent localizations can be seen with the products $w_{\upar}\chi_{\upar} \approx 0.300\approx 1/2\sqrt{\pi},$ as expected for the harmonic oscillator and $w_{\dar}\chi_{\dar} \approx 0.343 \approx 1/2\sqrt{2},$ corresponding to the exponential localization. 

With the knowledge of the induced localization for harmonic oscillator, we can understand the behavior of the condensate in the quasiperiodic potential. At large and moderate $\Omega_{0}$ it is very similar to the behavior in the harmonic oscillator potential with the frequency $\omega_{0}.$ With the decrease in  $\Omega_{0}$ the spin-down component spreads as $\chi_{\dar}^{-1}\sim 1/\sqrt{|\mu|}\sim \omega_{0}^{3/2}/\Omega_{0}^{2}$ and at $\Omega_{0}^{[{\rm cr}]}$ corresponding to $\chi_{\dar}x_{1}\lesssim 1$ extends to the wing minima causing the extension of the spin-up component followed by a fast increase in the width of both. Thus, from Eq. \eqref{eq:Vprimes} we can see that at the rescaling of the quasiperiodic potential as $\left(\tilde{V}_{1},\tilde{V}_{2}\right)=\nu\left({V}_{1},{V}_{2}\right)$ and  $\left(\tilde{k}_{1},\tilde{k}_{2}\right)=\kappa\left({k}_{1},{k}_{2}\right),$ the critical $\tilde{\Omega}_{0}^{[{\rm cr}]}$ rescales as $\nu^{-3/8}\kappa^{-1/4}.$
At a very small $\Omega_{0}$ the width follows that for free particle localized in a one-dimensional box. 

Now we discuss the role of the self-interaction in different spin components. 
Following Eq. \eqref{eq:Eint}, for this purpose we compare the quantities $N_{\upar}^{2}\chi_{\upar}\sim \Omega_{0}^{8}\lambda^{-15/2}$ and $\chi_{\dar}\sim \sqrt{|\mu|}\sim\Omega_{0}^{2}\lambda^{-3/2}.$ Thus, in weak Rabi fields, self interaction is much stronger in the spin-down component than in the spin up one since the spin up one has a low occupation probability. Thus, while the spin-up component interacts with the external potential, the spin-down component holds the self-interaction. A more detailed comparison of $g\chi_{\dar}$ and $|\mu|$ shows that self-interactions play an essential destructive role in the induced localization at $g\sim \sqrt{|\mu|}\sim 4\sqrt{\pi}\Omega_{0}^{2}\lambda^{-3/2},$ and this estimated is valid for other types of self-interaction with $g=g_{\upar\upar}=g_{\dar\dar}=g_{\upar\dar}=g_{\dar\upar}$ or    
Following Eq. \eqref{eq:Eint}, for this purpose we compare the quantities $N_{\upar}^{2}\chi_{\upar}\sim \Omega_{0}^{8}\lambda^{-15/2}$ and $\chi_{\dar}\sim \sqrt{|\mu|}\sim\Omega_{0}^{2}\lambda^{-3/2}.$ Thus, in weak Rabi fields, self interaction is much stronger in the spin-down component than in the spin up one since the spin up one has a low occupation probability. Thus, while the spin-up component interacts with the external potential, the spin-down component holds the self-interaction. A more detailed comparison of $g\chi_{\dar}$ and $|\mu|$ shows that self-interactions play an essential destructive role in the induced localization at $g\sim \sqrt{|\mu|}\sim 4\sqrt{\pi}\Omega_{0}^{2}\lambda^{-3/2},$ and this estimated is valid for other types of self-interaction with $g=g_{\upar\upar}=g_{\dar\dar}=g_{\upar\dar}=g_{\dar\upar}$ or    
Following Eq. \eqref{eq:Eint}, for this purpose we compare the quantities $N_{\upar}^{2}\chi_{\upar}\sim \Omega_{0}^{8}\lambda^{-15/2}$ and $\chi_{\dar}\sim \sqrt{|\mu|}\sim\Omega_{0}^{2}\lambda^{-3/2}.$ Thus, in weak Rabi fields, self interaction is much stronger in the spin-down component than in the spin up one since the spin up one has a low occupation probability. Thus, while the spin-up component interacts with the external potential, the spin-down component holds the self-interaction. A more detailed comparison of $g\chi_{\dar}$ and $|\mu|$ shows that self-interactions play an essential destructive role in the induced localization at $g\sim \sqrt{|\mu|}\sim 4\sqrt{\pi}\Omega_{0}^{2}\lambda^{-3/2},$ and this estimate is valid for other types of self-interaction with $g=g_{\upar\upar}=g_{\dar\dar}=g_{\upar\dar}=g_{\dar\upar}$ or $g=g_{\upar\dar}=g_{\dar\upar},g_{\upar\upar}=g_{\dar\dar}=0.$    
%%%%%%%%%%%%%%%%%%%%%%%%%%%%%%%%%%%%%%%%

\section{Resonant frequency for delocalization}
\begin{figure}[!htp]
    \centering
     \includegraphics[width=\linewidth]{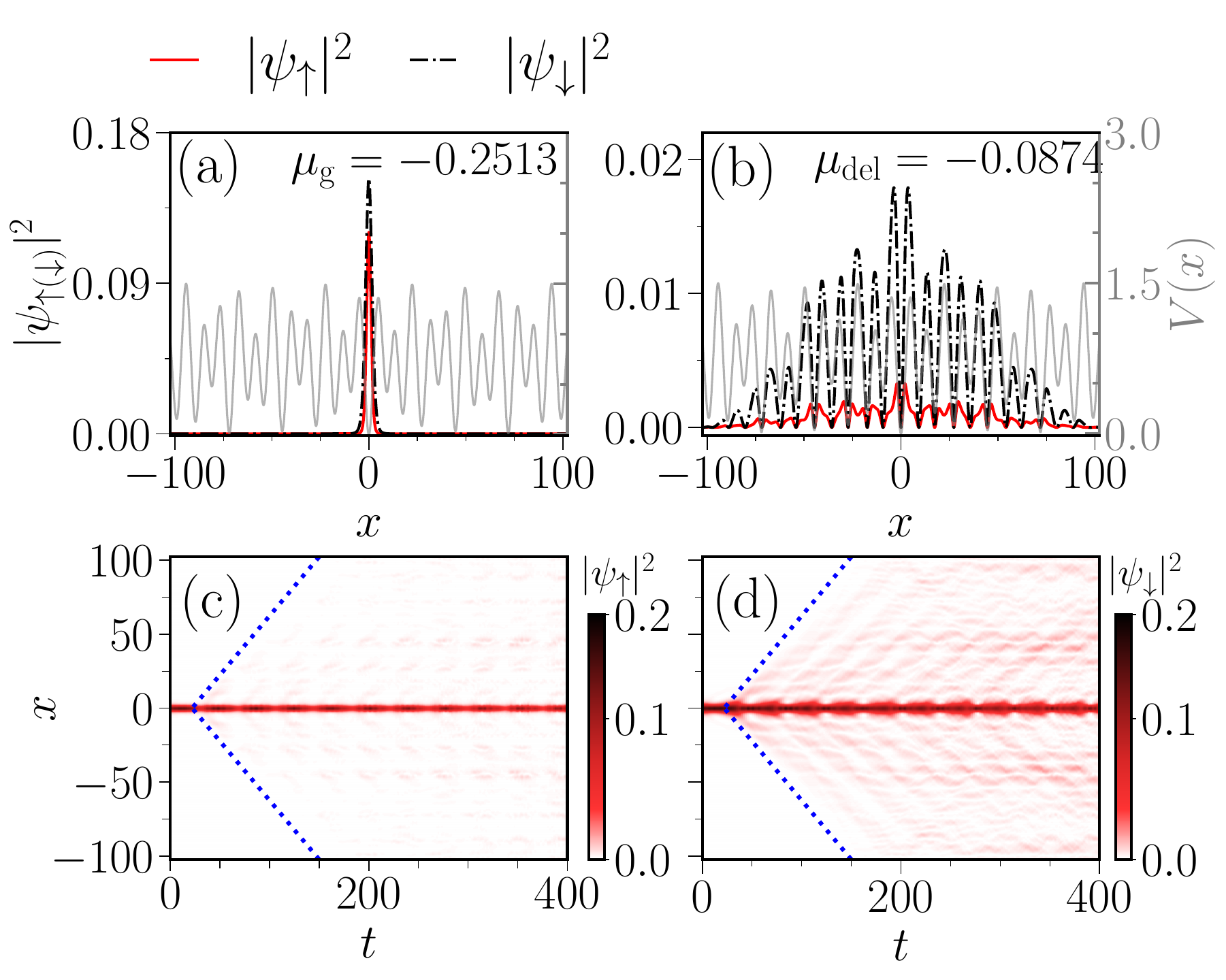}
    \caption{(upper panel) The condensate density $|\psi_{\upar (\dar)}|^2$ associated with the ground state (a) and the excited state (b) of the coupled GPEs. (lower panel) Pseudo-colormap representation of the condensate densities (c) $|\psi_{\upar}|^2$  and (d) $|\psi_{\dar}|^2$ at $\Omega_0 = 0.4, \Omega_1 = 0.2, \omega_{\rm osc} = 0.1639$. Here, the oscillation frequency $\omega_{\rm osc} = 0.1639$ is obtained by taking the energy difference between the ground (a) and excited states (b) of the system as $\omega_{\rm osc} \approx |\mu_g - \mu_{\rm ex}|\approx 0.1639$. Other parameters are $V_2/V_1 = 0.5, V_1 = 1.0, g = 0$.}
    \label{fig:res-freq-cal}
\end{figure}
Here we discuss estimation of the resonant frequency for non-interacting BECs using the matrix method. The solution of coupled GPEs (\ref{stat1(a)}-\ref{stat1(b)}) computes eigenvalues associated with the ground and excited states of the Hamiltonian, where the lowest eigenvalue corresponds to the ground state [see Fig.~\ref{fig:N-w-var-omega}(d)]. Here we utilize the difference between the ground-state and excited-state eigenvalues to obtain the approximate resonant frequency that drives the condensate delocalization.

We begin by solving the coupled GPE with quasiperiodic trap subjected to spin-up component by keeping $\Omega_0 = 0.4$. Subsequently, first fifty eigenvalues are obtained, in which the $0^{\rm th}$ eigenvalue $\mu_{g} \equiv \mu^{[0]} = -0.2513$ represents the ground state chemical potential [see Fig.~\ref{fig:res-freq-cal}(a)]. Following that, we examine excited states eigenvalues (~$\mu^1_{\rm ex}, \mu^2_{\rm ex}, \mu^{3}_{\rm ex} \ldots \mu^{50}_{\rm ex}$~) and wavefunction in order to find the appropriate delocalized condensate.
We found that for low-energy eigenstates, both  components are localized symmetrically on either side of $x = 0$, occupying different minima of the potential. For higher excited states, where $|\mu|$ is close to zero, the eigenstates are extended all over the space, overcoming the effect of the potential. We have chosen those eigenstates as the delocalized states. For example, in Fig.~\ref{fig:res-freq-cal}(b), we present the condensate density with eigenvalue $\mu_{\rm del} =  \mu^{23}_{\rm ex} = -0.0874$, where the density efficiently tunnels through the quasiperiodic trap. Therefore, the energy difference between the $\mu_g$ and $\mu_{\rm del}$ can be defined as the energy required to efficiently tunnel the condensate from the central minimum. Therefore, the approximate resonant frequency turns out to be $\omega_{\rm osc} \approx |\mu_g - \mu_{\rm del}|\approx 0.1639$ for $\Omega_0 = 0.4.$

Next, in Fig.~\ref{fig:res-freq-cal}(c) and ~\ref{fig:res-freq-cal}(d), we show the evolution of condensate density $|\psi_{\upar}|^2$, and $|\psi_{\dar}|^2$, respectively by keeping $\Omega_0 = 0.4, \Omega_1 = 0.2,$ and oscillation frequency at $\omega_{\rm osc} = 0.1639.$ Under this periodic driving, the condensate exhibits uniform expansion, albeit with a lower intensity compared to the case of $\omega_{\rm osc} = 0.2$ shown in Fig.~\ref{fig:t-x-density-non-int}(a2,b2). This suggests that $\omega_{\rm osc} \approx 0.1639$ resonantly drives the delocalization. It is important to note that, although the resonant frequency is estimated here by taking $\Omega_0 = 0.4$, the $\Omega_1$ term in $\Omega(t)$ make the dynamics nonlinear and, thus, effectively modifies the resonant frequency for delocalization.
\begin{figure}[!htp]
    \centering
    \includegraphics[width=0.85\linewidth]{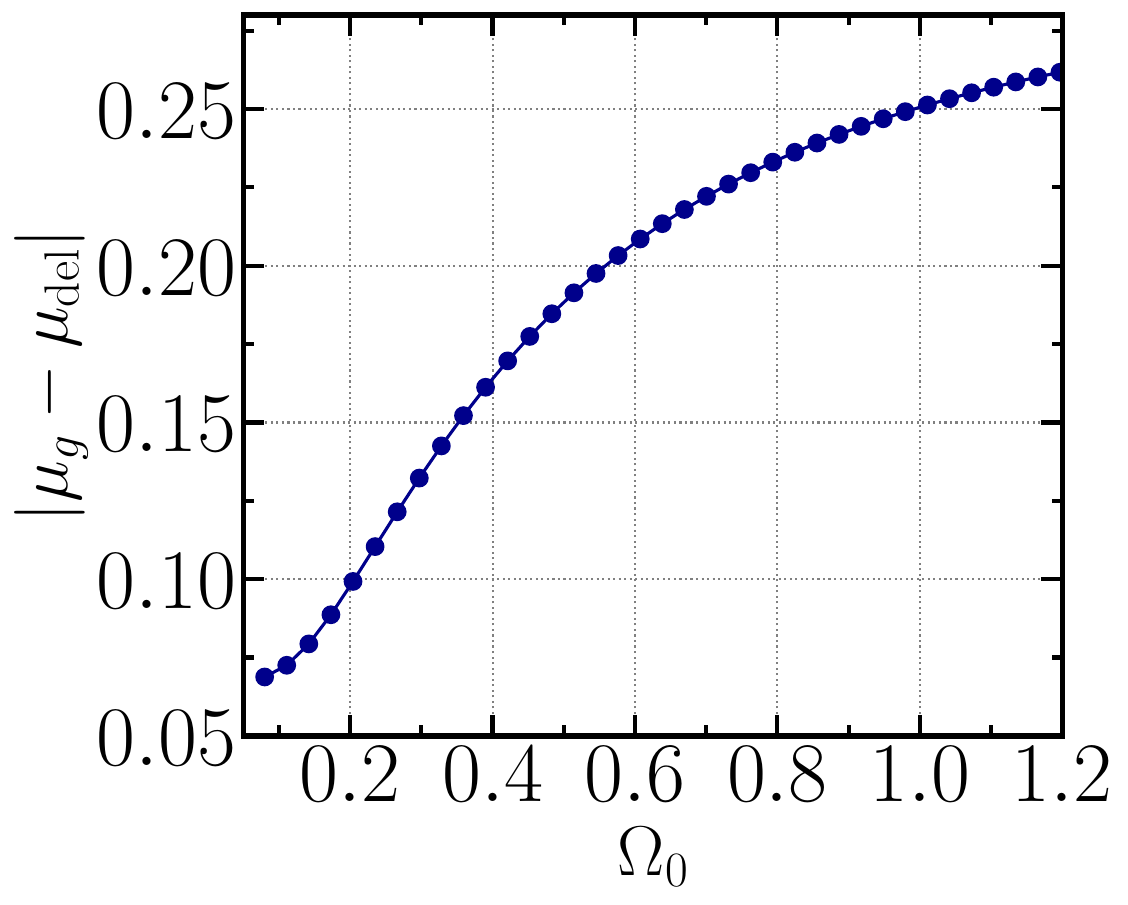}
    \caption{ Variation of the resonant frequency $\omega_{\rm osc}$ obtained by calculating the difference as $|\mu_g - \mu_{\rm del}|$ with Rabi coupling $\Omega_0$ by keeping $V_2/V_1 = 0.5$, $V_1 = 1.0$, and $g = 0$. The resonant frequency initially increases almost linearly with $\Omega_0$ and then saturates at higher values of $\Omega_0.$}
    \label{fig:res-freq-Omega0}
\end{figure}

In Fig.~\ref{fig:res-freq-Omega0}, we illustrate the variation of the resonant frequency as a function of $\Omega_0$. For low values of $\Omega_0$, the resonant frequency increases almost linearly with $\Omega_0$, followed by a saturation trend at larger values, where in the ground state the components behave as Gaussian localized states strongly localized in harmonic confinement with $N_{\upar}\approx N_{\dar}.$ Under such conditions, the potential-free spin-down component closely follows the spin-up component. Although the matrix method gives only an approximate value of the resonant frequency, it provides valuable insights into the Rabi-driven delocalization of a condensate.

\section{Effect of Rabi coupling on spin polarization}
\label{appen:C}
{Here we demonstrate the role of Rabi coupling in formation of spin polarization of the BEC. The spin polarization can be defined by attributing spin-dependent observables as:
\begin{eqnarray}\label{eq:spin}
&&\langle\sigma_{x}\rangle=\int_{-\infty}^{\infty} (\psi_{\upar}^{\dagger}\psi_{\dar} + \psi_{\dar}^{\dagger}\psi_{\upar}) dx, \\
&&\langle\sigma_{y}\rangle= {\rm i} \int_{-\infty}^{\infty} (\psi_{\upar}^{\dagger}\psi_{\dar} - \psi_{\dar}^{\dagger}\psi_{\upar}) dx,\\
&&\langle\sigma_{z}\rangle= \int_{-\infty}^{\infty} (|\psi_{\upar}|^2 - |\psi_{\dar}|^2)dx.
\end{eqnarray}
\begin{figure}
    \centering
    \includegraphics[width=\linewidth]{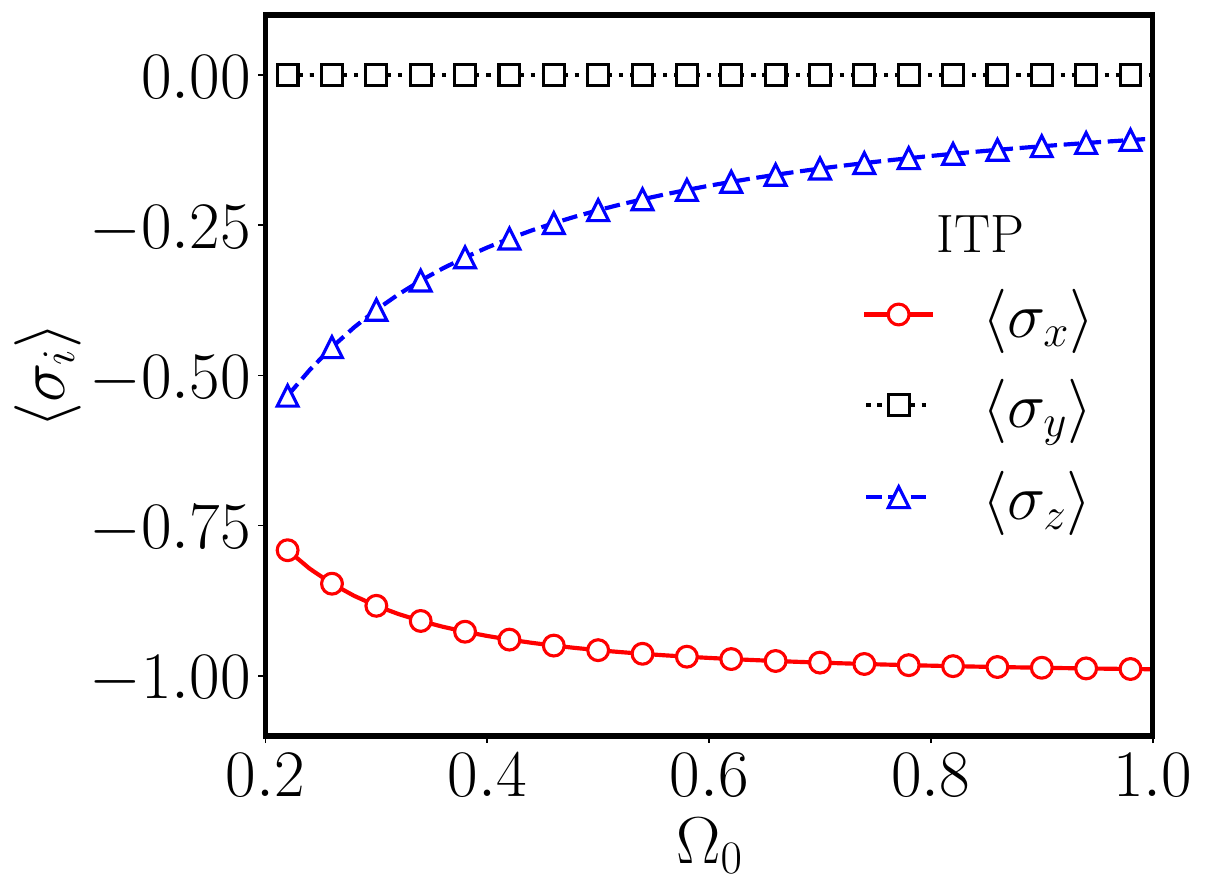}
    \caption{Variation of $\langle \sigma_x \rangle$, $\langle \sigma_y \rangle,$ and $\langle \sigma_z \rangle$ with Rabi coupling $\Omega_0$. The variation of $\langle \sigma_x \rangle$, and $\langle \sigma_y \rangle$ demonstrates the change in the spin polarization with the variation of $\Omega_0$. The other parameters are $V_2/V_1 = 0.5, V_1 = 1.0, g = 0.$}
    \label{fig:spin-observables}
\end{figure}
Figure~\ref{fig:spin-observables} shows the variation of the ground state spin-related observables $\langle \sigma_i \rangle$ as a function of $\Omega_0$. As expected, with the increase in $\Omega_0$, $\langle \sigma_x \rangle$ tends to $-1$, and $\langle \sigma_z \rangle$ characterizing the spin polarization ($\langle \sigma_z \rangle = N_{\upar} - N_{\dar}$) of the system approaches zero as the spin components balance each other with population exchange. Such modifications of spin polarization are well-known in multicomponent BECs with unequal self-interactions. Here, however, we demonstrate that within this spin-asymmetric potential model, the spin polarization $\langle\sigma_{i}\rangle-$components can be tuned effectively even in a linear condensate.}

\bibliography{references.bib}

\end{document}